\providecommand{\solarprop}{\texttt{SOLAR\-PROP}}
\providecommand{\solarproptwo}{\texttt{SOLAR\-PROP~2.0}}

\documentclass[final,3p,times,nopreprintline,12pt]{elsarticle}

\usepackage{hyperref}
\usepackage{listings}
\usepackage{siunitx}

\usepackage{xcolor}

\definecolor{codegreen}{rgb}{0,0.6,0}
\definecolor{codegray}{rgb}{0.5,0.5,0.5}
\definecolor{codepurple}{rgb}{0.58,0,0.82}
\definecolor{backcolour}{rgb}{0.95,0.95,0.92}

\lstdefinestyle{mystyle}{
    backgroundcolor=\color{backcolour},   
    commentstyle=\color{codegreen},
    keywordstyle=\color{magenta},
    numberstyle=\tiny\color{codegray},
    stringstyle=\color{codepurple},
    basicstyle=\ttfamily\footnotesize,
    breakatwhitespace=false,         
    breaklines=true,                 
    captionpos=b,                    
    keepspaces=true,                 
    numbers=none,                    
    numbersep=5pt,                  
    showspaces=false,                
    showstringspaces=false,
    showtabs=false,                  
    tabsize=2
}

\usepackage{amssymb}
\usepackage{amsmath}

\usepackage{lineno}

\begin{document}

\begin{frontmatter}



\title{Solarprop 2.0: Modern charge-sign dependent solar modulation for everyone}


\author{Henning Gast} 

\affiliation{organization={I. Physikalisches Institut B, RWTH Aachen University},
            addressline={Sommerfeldstr. 14}, 
            city={52074 Aachen},
            country={Germany}}

\begin{abstract}
  I present a greatly improved version of \solarprop{}, an open source
  framework for building models of charge-sign dependent solar
  modulation of cosmic rays in the heliosphere. The transport equation
  of charged particles is solved by converting it to a set of
  stochastic differential equations (SDEs). The code now includes the
  full set of SDEs in two spatial dimensions, allowing for magnetic
  fields with latitudinal components, and it has been generalized to
  accommodate models in three spatial dimensions. Due to code
  parallelization and optimization, a speedup of up to three orders of
  magnitude with respect to the original version is achieved. Python
  bindings have been added for ease of use. The code has been
  validated against the results of several model calculations found in
  the literature. While excellent agreement is found in many cases,
  important differences are observed for two model
  implementations. Possible root causes for the differences have been
  investigated.
\end{abstract}

\begin{highlights}
\item Full set of 2D stochastic differential equations, allowing for
  magnetic fields with latitudinal components, beyond the Parker
  spiral.
\item Generalization to models in three spatial dimensions.
\item New models for validation and comparison to existing literature.
\item Massive speedup by OpenMP parallelization and code optimizations.
\item New Python interface for ease of use.
\item Modern build system.
\item System of units, visualization scripts, better handling of data files.
\item Several bug fixes.
\end{highlights}

\end{frontmatter}


\section{Introduction}
\label{sec:intro}
Charged cosmic rays are important messengers about our galactic
surroundings. Their composition, energy spectra and arrival directions
carry information about stellar evolution, interstellar matter and the
turbulent magnetic fields in the Galaxy. Protons, alpha particles,
electrons, and heavier nuclei make up the bulk of cosmic rays. Within
our galaxy, supernova remnants~\cite{Drury1994} are believed to be the
main sites where cosmic rays are accelerated. Instruments on
high-altitude balloons have measured cosmic rays for decades in the
last century. In recent years, space-based detectors such as
PAMELA~\cite{Picozza2007} provided data with unprecedented
accuracy. In particular, the Alpha Magnetic Spectrometer (AMS) on the
International Space Station has turned cosmic-ray physics into a field
of precision study~\cite{Aguilar2021}, covering the energy range from
the gigaelectronvolt (GeV) to the teraelectronvolt regime with
percent-level uncertainties. Its measurements also open a detailed
view at the more exotic components of cosmic rays, namely positrons
and antiprotons. Interactions of primary cosmic rays with interstellar
matter will lead to the production of a certain amount of these
antimatter particles. But their relative rareness makes them sensitive
probes of exciting new physics. For example, pairwise annihilations of
dark matter particles or additional astrophysical sources like pulsars
could modify the energy spectra and/or arrival directions of cosmic
rays in characteristic ways~\cite{Mauro2016}.

Measurements of cosmic rays are performed in the orbit of the Earth or
even in its upper atmosphere, with the exception of the Voyager probes
that are now traveling through interstellar
space~\cite{Stone2019}. But on the last steps of their journey towards
Earth, the energy spectra and arrival directions of cosmic rays
undergo dramatic changes when the charged particles interact with the
magnetized solar wind inside the heliosphere. This phenomenon is
called solar modulation~\cite{Potgieter2013a}. The magnitude of the
effects caused by solar modulation varies with time due to changes in
solar activity over the 11-year solar cycles and the reversal of the
polarity of the heliospheric magnetic field in each cycle. Any model
of cosmic-ray origin and propagation has to take the effects of solar
modulation into account before conclusions can be drawn. Even though
the amplitude of solar modulation decreases with energy, it is still
significant up to tens of GeV compared to the small uncertainties
provided by detectors like AMS. Even the modeling of phenomena at
high energies can be affected by solar modulation, for example, when
parameters describing backgrounds from secondary production are
determined from fits to the low-energy data~\cite{Aguilar2019}. Even
more acutely, some models of dark matter annihilation predict an
excess in the cosmic-ray antiproton spectrum at a few GeV, thus in an
energy range strongly affected by solar
modulation~\cite{Cuoco2017}. In essence, one has to understand solar
modulation in order to extract the local interstellar spectra of
cosmic rays that carry information about our cosmic surroundings, and
potentially new physics, from measurements taken deep inside the
heliosphere, at Earth.

A proper description of solar modulation is an important aspect in any
model for cosmic-ray origin and propagation. At the same time, solar
modulation tends to be treated rather superficially in many
models. This is in part due to the existence of the force-field
approximation~\cite{Gleeson1968}, which gives a first-order
description of modulation effects in the form of a simple analytic
formula. But it is also due to a lack of suitable freely available
software allowing for a realistic treatment of the physics behind
solar modulation. Such software needs to be \emph{fast}, because it
will often be used in high-dimensional parameter optimizations in
conjunction with cosmic-ray propagation models such as
GALPROP~\cite{Strong1998} or DRAGON2~\cite{Evoli2017};
\emph{accessible}, so that it is easily interfaced with models
predicting local interstellar spectra of cosmic-ray particles;
\emph{transparent}, so that the details of the implementation of the
solar modulation models can be inspected and the reproducibility of
results can be ensured; and \emph{versatile}, so that solar modulation
models can be easily modified or new models implemented.

With similar goals in mind, the \solarprop{} code was developed and
published by Rolf Kappl~\cite{Kappl2015} in~2016, as an open source
tool for charge-sign dependent solar modulation of cosmic rays. The
code was validated by implementing four different previously published
models and comparing the modulated energy spectra to results shown in
the respective publications. \solarprop{} solves the transport
equation by converting it to a set of stochastic differential
equations (SDEs) which are then evaluated with Monte Carlo
methods~\cite{Zhang1999,Kopp2012}.

I present \solarproptwo{}, a greatly improved version of
\solarprop{}. The main improvements are summarized as follows. 1) The
full set of two-dimensional stochastic differential equations has been
implemented, allowing for non-Parkerian magnetic fields with
latitudinal components. In addition, models with Parkerian magnetic
fields can now be treated in all three spatial dimensions. 2) A change
of the coordinate system, i.e.~using $\cos\theta$ instead of the
colatitude $\theta$ directly, together with several minor bug fixes
lead to increased numerical stability and better agreement of the
calculated spectra with published results. 3) The run time of the code
has been massively reduced, first by avoiding unnecessary and
duplicated calculations, and second by parallelizing the evaluation of
the SDEs using OpenMP. The first effect saves roughly a factor of 10,
depending on the details of the model, and the second effect reduces
the run time approximately by the number of computing cores used for
the calculations. So on a modern workstation with $\sim\!100$ cores,
the speedup with respect to vanilla \solarprop{} can amount to three
orders of magnitude.

While the core structure of the original \solarprop{} was retained, most
of the code was rewritten. The model implementation and solution of
the transport equation are still written in C++, but the model can now
be configured and the results of the calculation obtained using Python
bindings. The installation of \solarproptwo{} proceeds via a single
\texttt{pip} command. These improvements make \solarproptwo{} accessible
for models of cosmic-ray origin and propagation. A simple usage
example looks like this:

\begin{lstlisting}[language=Python, caption=Usage example for the Python interface.]
import numpy as np
import solarprop

# configure model and its parameters
parameters = {'model': 'BP1989',
              'polarity': 1,
              'mass': 0.93827231,
              'charge': 1}

# set kinetic energy binning (in units of GeV)
ekin = np.geomspace(0.02, 25.0, num=25)

# define local interstellar spectrum (LIS)
LIS = lambda T, m: np.sqrt(T**2 + 2*T*m) * np.pow(T + m, -3.6)

# create solarprop model
s = solarprop.Solarprop(parameters)

# calculate modulated flux at points given by "ekin" array
flux_mod = s.modulate(ekin, LIS(ekin, parameters['mass']))
\end{lstlisting}

The scope of this paper covers the development and validation of
\solarproptwo{} as a general, \emph{open source} framework for the
implementation of models of charge-sign dependent solar
modulation. Hopefully, this development will facilitate the next step,
namely describing the high-quality time-resolved data on various
cosmic-ray species now available, and thus constraining the local
interstellar spectra of cosmic rays, with an \emph{open source} model
of solar modulation. Other open source codes for solar modulation of
cosmic rays have been published, such as
Geliosphere~\cite{Solanik2023} and
COSMICA~\cite{DellaTorre2025,Bacciu2026,DellaTorre2026}. These two
rely on graphics processing units (GPUs) to speed up the calculations,
which makes them arguably less accessible than the simple CPU-based
approach followed by \solarproptwo{}.

The source code, documentation, and examples of \solarproptwo{} are
available
online\footnote{\url{https://github.com/henningrwth/solarprop2}}. A
detailed summary of the changes and improvements with respect to
vanilla \solarprop{} is found in \ref{sec:list-changes}.

\section{Particle transport in the heliosphere}
\label{sec:physics}
The
\solarproptwo{} code solves the transport equation~\cite{Parker1965}
\begin{equation}
  \label{eq:parker}
  \frac{\partial f}{\partial t}=\vec{\nabla}\cdot\left(\kappa\cdot\vec{\nabla}f-\vec{V}f\right)+\frac{1}{3p^2}(\vec{\nabla}\cdot\vec{V}_{sw})\frac{\partial p^3f}{\partial p}
\end{equation}
for the propagation of cosmic rays in the heliosphere. Here the
distribution of a given cosmic-ray species in phase space is described
by the omnidirectional distribution function $f(\vec{x}, p)$, where
$p$ is particle momentum, such that the differential particle flux is
given by $\Phi=p^2f$. In the transport equation, $\kappa$ is the
symmetric diffusion tensor, $\vec{V}_{sw}$ is the solar wind speed,
$\vec{V}=\vec{V}_{sw}+\vec{V}_{dr}$, and $\vec{V}_{dr}$ is the
pitch-angle averaged guiding center drift velocity in the heliospheric
magnetic field. The transport equation describes convection by the
expanding solar wind, gradient, curvature, and current sheet drifts in
the heliospheric magnetic field, diffusion, and adiabatic energy
changes.

The transport equation is solved by expressing all terms in spherical
heliocentric coordinates $(r,\theta,\phi)$ and bringing it to the
standard form of the time backward Kolmogorov
equation~\cite{Strauss2017},
\begin{equation}
  \label{eq:kolmogorov}
  \frac{\partial f}{\partial t}=\sum\limits_iA_i(\vec{q},t)\frac{\partial f}{\partial q_i}+\frac{1}{2}\sum\limits_{i,j}\left[B(\vec{q},t)B^T(\vec{q},t)\right]_{i,j}\frac{\partial^2 f}{\partial q_i\partial q_j}.
\end{equation}
The corresponding set of stochastic differential equations is then
given by
\begin{equation}
  \label{eq:sdeq}
  \mathrm{d}\vec{q}=A(\vec{q},t)\,\mathrm{d}t+B(\vec{q},t)\,\mathrm{d}W,
\end{equation}
where $\mathrm{d}W$ is the increment of the Wiener process.

Two derive the set of SDEs for the two-dimensional case, the azimuthal
coordinate is neglected and $\vec{q}=(r,\mu,p)$, with
$\mu=\cos\theta$, is chosen as the set of coordinates. Assuming that
the solar wind is radially directed, $\vec{V}_{sw}=V_{sw}\,\vec{e}_r$,
defining
$\vec{V}_{dr}\equiv V_{dr}\,\vec{e}_r + V_{d\theta}\,\vec{e}_\theta$,
and using the fact that $\vec{\nabla}\cdot\vec{V}_{dr}=0$ leads to the
following set of SDEs (the full derivation is shown in
\ref{app:sdes2d}):
\begin{equation}
  \label{eq:sdes2d}
  \begin{aligned}
    \mathrm{d}r
    &= \left[\frac{1}{r^2}\frac{\partial}{\partial r}(r^2\kappa_{rr}) -
      \frac{1}{r}\frac{\partial}{\partial\mu}(\kappa_{r\theta}\sqrt{1-\mu^2})
      - V_{sw}-V_{dr}\right]\,\mathrm{d}t \\
    &+ w_r\,\frac{\sqrt{- 2 \kappa_{r \theta}^{2} + 2 \kappa_{rr}\kappa_{\theta \theta}}}{\sqrt{\kappa_{\theta \theta}}}\,\sqrt{\mathrm{d}t} - w_\mu \frac{\sqrt{2} \kappa_{r \theta}}{\sqrt{\kappa_{\theta \theta}}} \,\sqrt{\mathrm{d}t},\\
    \mathrm{d}\mu &= \left[
      -\frac{\sqrt{1-\mu^2}}{r^2}\frac{\partial}{\partial
        r}(r\kappa_{r\theta}) +
      \frac{1}{r^2}\frac{\partial}{\partial\mu}(\kappa_{\theta\theta}(1-\mu^2))+\frac{\sqrt{1-\mu^2}}{r}V_{d\theta}
    \right]\,\mathrm{d}t \\
    &+ w_\mu \frac{\sqrt{2(1 - \mu^{2})\kappa_{\theta \theta}}}{r} \,\sqrt{\mathrm{d}t},\\
    \mathrm{d}p & =\frac{p}{3r^2}\frac{\partial r^2V_{sw}}{\partial r}\,\mathrm{d}t.\\
  \end{aligned}
\end{equation}
Here, $w_r$ and $w_\mu$ are Gaussian random numbers with a mean of
zero and unit variance, and $\mathrm{d}t$ is the time increment. The
elements of the diffusion tensor in spherical heliospheric coordinates
are denoted by $\kappa_{rr}$, $\kappa_{\theta\theta}$, and
$\kappa_{r\theta}$. Throughout \solarproptwo, the SDEs are integrated
using the computationally efficient backward-in-time
approach~\cite{Bobik2016}. In each kinetic energy bin $T_i$, a large
number ($\mathcal{O}(1000)$) of pseudo-particles is created at the
location of the Earth. Each pseudo-particle is propagated through the
heliosphere, by adjusting its phase space coordinates $(r,\mu,p)$
according to the SDEs in each time step, until it reaches the border
of the heliosphere. The randomness inherent in the diffusive particle
transport in the heliosphere is reflected in the Gaussian random
numbers $w_r$ and $w_\mu$ that will cause each pseudo-particle
trajectory to be different. The solution of the transport equation
then arises from the ensemble average of the
pseudo-particles~\cite{Pei2010}. From the pseudo-particle states at
the heliopause, the probability matrix $G(T_i,T_j)$ is extracted,
containing the probabilities for a particle in kinetic energy bin
$T_i$ at Earth to originate from the kinetic energy bin $T_j$ at the
heliopause. The given local interstellar spectrum (LIS)
$\Phi_\mathrm{LIS}(T)$, i.e. the differential flux of particles as a
function of kinetic energy $T$ at the heliopause, then provides the
boundary condition for the solution. The modulated flux at Earth is
$\Phi_\mathrm{mod}(T)=p^2(T)f_\mathrm{mod}(T)$, with the phase-space
density calculated as
\begin{equation}
  \label{eq:modflux}
  f_\mathrm{mod}(T_i)=\sum\limits_{j}G(T_i,T_j)f_\mathrm{LIS}(T_j),
\end{equation}
with $f_\mathrm{LIS}(T)=\Phi_\mathrm{LIS}(T)/p^2(T)$, and
$p^2(T)=T^2+2mT$, where $m$ is the particle mass.

Instead of using $\theta$ directly like in vanilla \solarprop{},
expressing the SDEs in terms of $\mu=\cos\theta$ is beneficial for the
numerical stability of the solution because it prevents large jumps in
$\theta$ in the polar regions of the heliosphere in certain
models. Using momentum $p$ instead of kinetic energy $T$ as the
kinematic variable simplifies the equations because no additional loss
term appears in this case~\cite{Bobik2016}. (The loss term was
neglected in vanilla \solarprop{}).\newline
\par
An alternative set of SDEs has been derived and implemented for models
that treat the particle propagation in all three spatial dimensions
but without polar diffusion,
$\kappa_{r\theta}=\kappa_{\theta{}r}=\kappa_{\theta\phi}=\kappa_{\phi\theta}=0$. In
this case, the SDEs are given by (full derivation
in~\ref{app:sdes3d}):
\begin{equation}
  \label{eq:sdes3d}
  \begin{aligned}
    \mathrm{d}r&=\left[\frac{1}{r^2}\frac{\partial}{\partial r}(r^2\kappa_{rr}) + \frac{1}{r\sqrt{1-\mu^2}} \frac{\partial\kappa_{r\phi}}{\partial\phi} - V_{sw} - V_{dr}\right]\,\mathrm{d}t\\
    &+w_r\, \sqrt{2\kappa_{rr}-\frac{2\kappa_{r\phi}^2}{\kappa_{\phi\phi}}} \,\sqrt{\mathrm{d}t} + w_\phi \frac{\sqrt{2} \kappa_{r \phi}}{\sqrt{\kappa_{\phi\phi}}} \,\sqrt{\mathrm{d}t},\\
    \mathrm{d}\mu&=\left[ \frac{1}{r^2}\frac{\partial}{\partial\mu}(\kappa_{\theta\theta}(1-\mu^2))+\frac{\sqrt{1-\mu^2}V_{d\theta}}{r} \right]\,\mathrm{d}t\\
    &+w_\mu \frac{\sqrt{2(1-\mu^2)\kappa_{\theta\theta}}}{r} \,\sqrt{\mathrm{d}t},\\
    \mathrm{d}\phi&=\left[ \frac{1}{r^2\sqrt{1-\mu^2}}\frac{\partial(r\kappa_{r\phi})}{\partial{}r} + \frac{1}{r^2(1-\mu^2)} \frac{\partial\kappa_{\phi\phi}}{\partial\phi} -\frac{V_{d\phi}}{r\sqrt{1-\mu^2}} \right]\,\mathrm{d}t\\
    &+w_\phi \frac{\sqrt{2\kappa_{\phi\phi}}}{r\sqrt{1-\mu^2}} \,\sqrt{\mathrm{d}t},\\
    \mathrm{d}p&=\frac{p}{3r^2}\frac{\partial r^2V_{sw}}{\partial r}\,\mathrm{d}t.\\
  \end{aligned}
\end{equation}
Here, $w_r$, $w_\mu$, and $w_\phi$ are Gaussian random numbers with a mean of
zero and unit variance.

\section{Validation}
\label{sec:validation}
\subsection{Update on models from vanilla \solarprop{}}
The \solarproptwo{} code has been validated against the same four
solar modulation models for cosmic-ray protons found in the literature
as used for the first version of the software. For details on the
magnetic field configuration, diffusion coefficients, treatment of
gradient and curvature drifts as well as drifts along the heliospheric
current sheet, and the assumed local interstellar spectrum (LIS) in
each model, the reader is referred to the original
publication~\cite{Kappl2015} where the relevant formulae and parameter
values are given. The validation models are briefly summarized in the
following. The results calculated for them with \solarproptwo{} are
compared to those given in the respective publications and to the
results obtained with vanilla \solarprop{}.

\begin{figure}[tp]
  \centering
  \includegraphics[width=\textwidth]{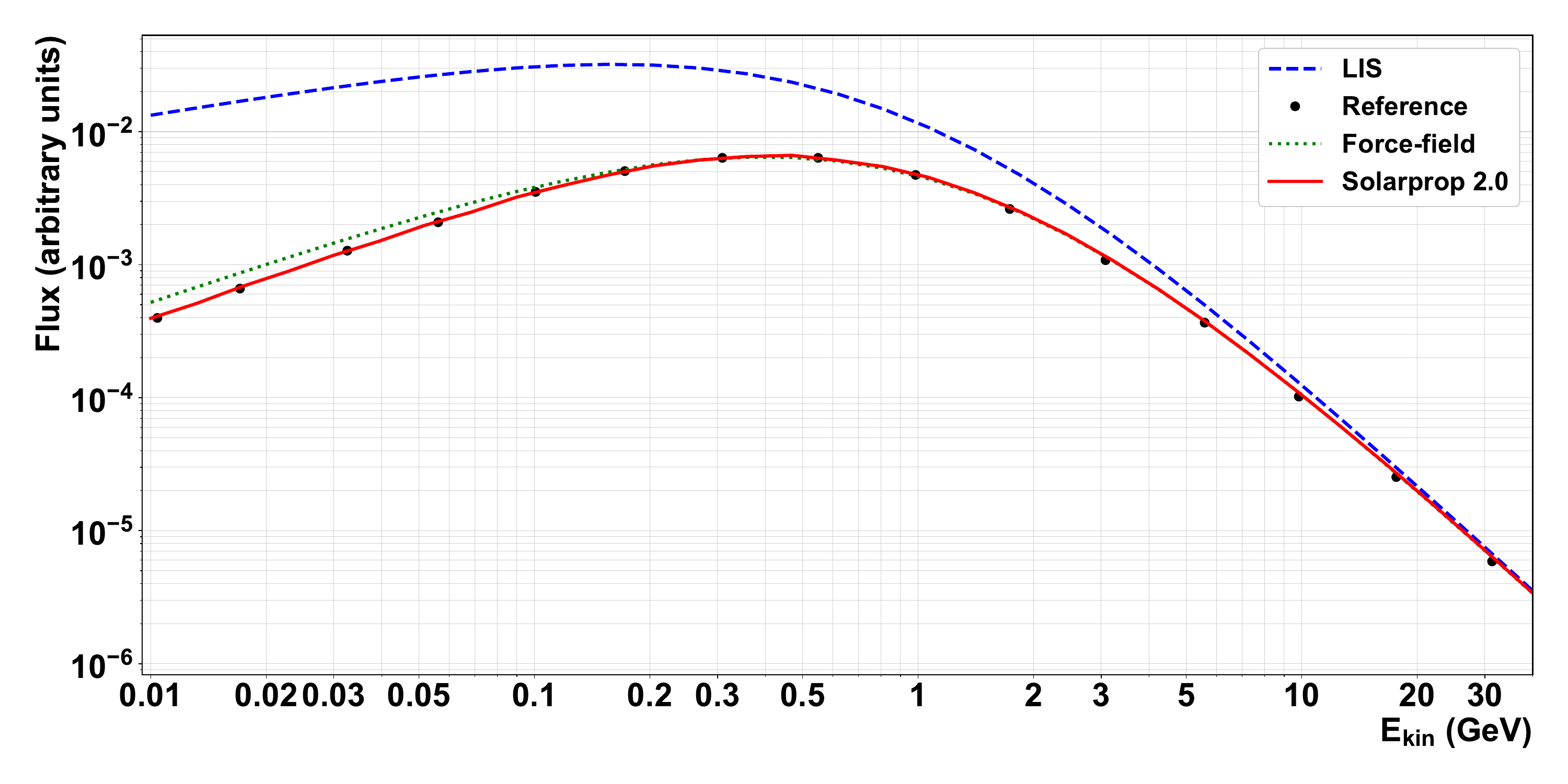}
  \caption{Validation of \solarproptwo{} with the model of Yamada et
    al.~\cite{Yamada1998}. The points labeled ``Reference'' have been digitized
    from the calculation results displayed in Figure~4 of Ref.~\cite{Yamada1998}.}
  \label{fig:yamada}
\end{figure}
\begin{figure}[tp]
  \centering
  \includegraphics[width=\textwidth]{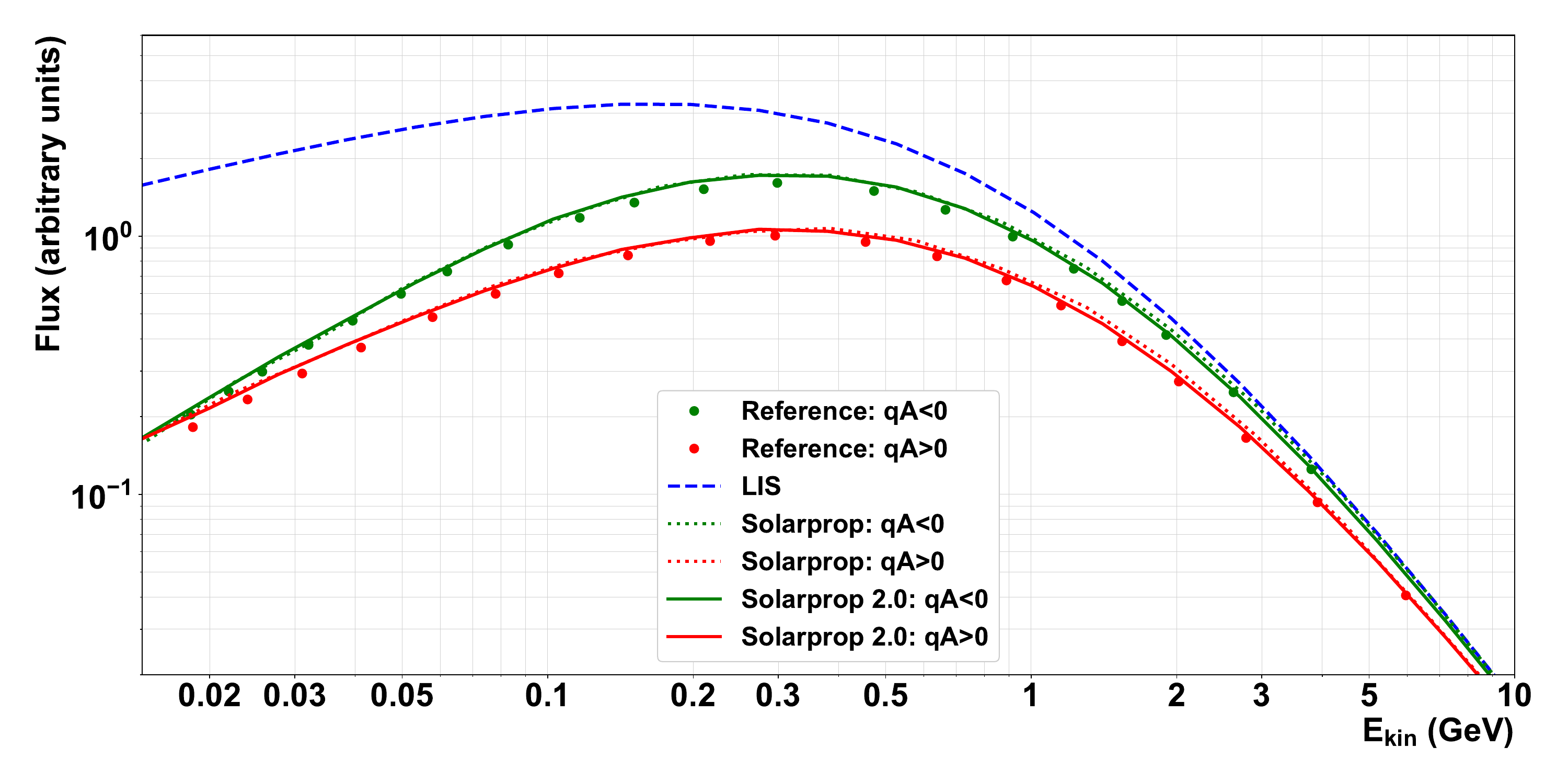}
  \caption{Validation of \solarproptwo{} with the model of Jokipii and
    Kopriva~\cite{Jokipii1979}. The points labeled ``Reference'' have
    been digitized from the calculation results displayed in Figure~2
    of Ref.~\cite{Jokipii1979}.}
  \label{fig:jk}
\end{figure}
The first model, published by Yamada et al.~\cite{Yamada1998}, is a
simple one-dimensional plain diffusion model with a scalar diffusion
coefficient $\kappa_0$ that is constant throughout the heliosphere. As
seen in Fig.~\ref{fig:yamada}, the result obtained with
\solarproptwo{} is still in excellent agreement with the published
result. The figure also demonstrates that this model is similar to the
analytic force-field approximation~\cite{Gleeson1968}, for a
modulation potential $\varphi = V_{sw}(R_h-\qty{1}{AU})/3\kappa_0$,
where $R_h=\qty{100}{AU}$ is the value adopted for the radius of the
heliosphere.

The other three validation models are two-dimensional and assume a
Parker spiral for the heliospheric magnetic field,
\begin{equation}
  \label{eq:parkerspiral}
  \vec{B}=AB_0\frac{r_0^2}{r^2}
  \left[\,\vec{e}_r-\Gamma\,\vec{e}_\phi\right] \times
\left[1-2H(\theta-\theta^\prime)\right]
\quad\text{with}\quad
  \Gamma=\frac{\Omega r\sin\theta}{V_{sw}},
\end{equation}
where $\Omega=2\pi/\qty{25.4}{days}$ is the average angular velocity
of the Sun, $H$ is the Heaviside step function, and $\theta^\prime$ is
the polar angle determining the location of the heliospheric current
sheet. The following simplified formula is used~\cite{Jokipii1981},
except where stated otherwise:
\begin{equation}
  \label{eq:location_hcs}
  \theta^\prime=\frac{\pi}{2}+\alpha\sin\left(\phi+\frac{\Omega}{V_{sw}}r\right),
\end{equation}
where $\alpha$ is the tilt angle of the current sheet. The magnitude
of the magnetic field is determined by the values of
$B_0=\qty{3.4}{\nano\tesla}$ and $r_0=\qty{1}{AU}$, and its polarity
is given by $A=\pm{}1$. In general, the drift velocity field is
calculated from the magnetic field by
\begin{equation}
  \label{eq:drift}
  \vec{V}_{dr}=\frac{pv}{3q}\vec{\nabla}\times\frac{\vec{B}}{B^2},
\end{equation}
where $v$ is the velocity of the pseudo-particle and $q$ is its
charge, and decomposed into a regular component $\vec{v}_\mathrm{gc}$
and a term $\vec{v}_\mathrm{hcs}$ which formally arises from the
derivative of the Heaviside step function and is used to describe
drift along the heliospheric current sheet:
$\vec{V}_{dr}=\vec{v}_\mathrm{gc}+\vec{v}_\mathrm{hcs}$. In some
models, these terms are regularized by a transition function, the
detailed equations are given in Ref.~\cite{Kappl2015}.

The second model, published by Jokipii and Kopriva~\cite{Jokipii1979},
uses a position-independent diffusion tensor and a flat heliospheric
current sheet. The results for opposite polarities of the heliospheric
magnetic field obtained with \solarproptwo{} are still in good
agreement with the published ones, and there is a small improvement
with respect to the original \solarprop{} (Fig.~\ref{fig:jk}).

\begin{figure}[tp]
  \centering
  \includegraphics[width=\textwidth]{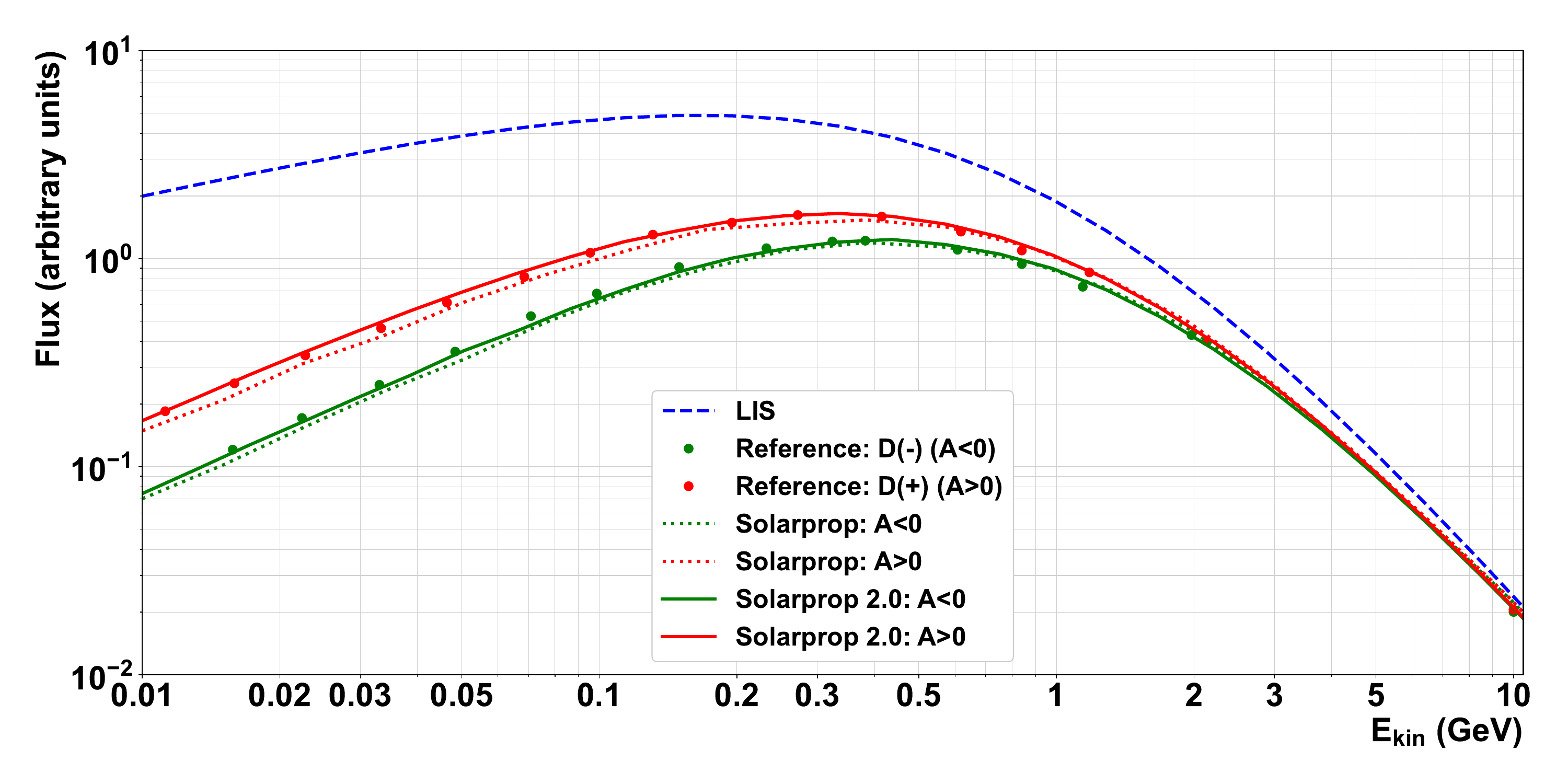}
  \caption{Validation of \solarproptwo{} with the model of Potgieter
    and Moraal~\cite{Potgieter1985}. The points labeled
    ``Reference'' have been digitized from the calculation results
    displayed in Figure~6 of Ref.~\cite{Potgieter1985}.
    (Unfortunately, the publication is only available online as a scan
    where the figure is heavily distorted above $\sim\!\qty{2}{GeV}$
    so no values were digitized in the affected region.)}
  \label{fig:pm}
\end{figure}
The third model, published by Potgieter and
Moraal~\cite{Potgieter1985}, combines diffusion and drift effects and
simulates a wavy heliospheric current sheet. The results for opposite
polarities of the heliospheric magnetic field obtained with
\solarproptwo{} are in excellent agreement with the published ones
(Fig.~\ref{fig:pm}). There is a visible improvement in the agreement
with respect to vanilla \solarprop{} for the $A>0$ case, where most of
the propagation happens in the polar regions of the heliosphere. This
area is most affected by the change from SDEs formulated in terms of
$\mathrm{d}\theta$ to those using $\mathrm{d}\!\cos\theta$.

\begin{figure}[tp]
  \centering
  \includegraphics[width=\textwidth]{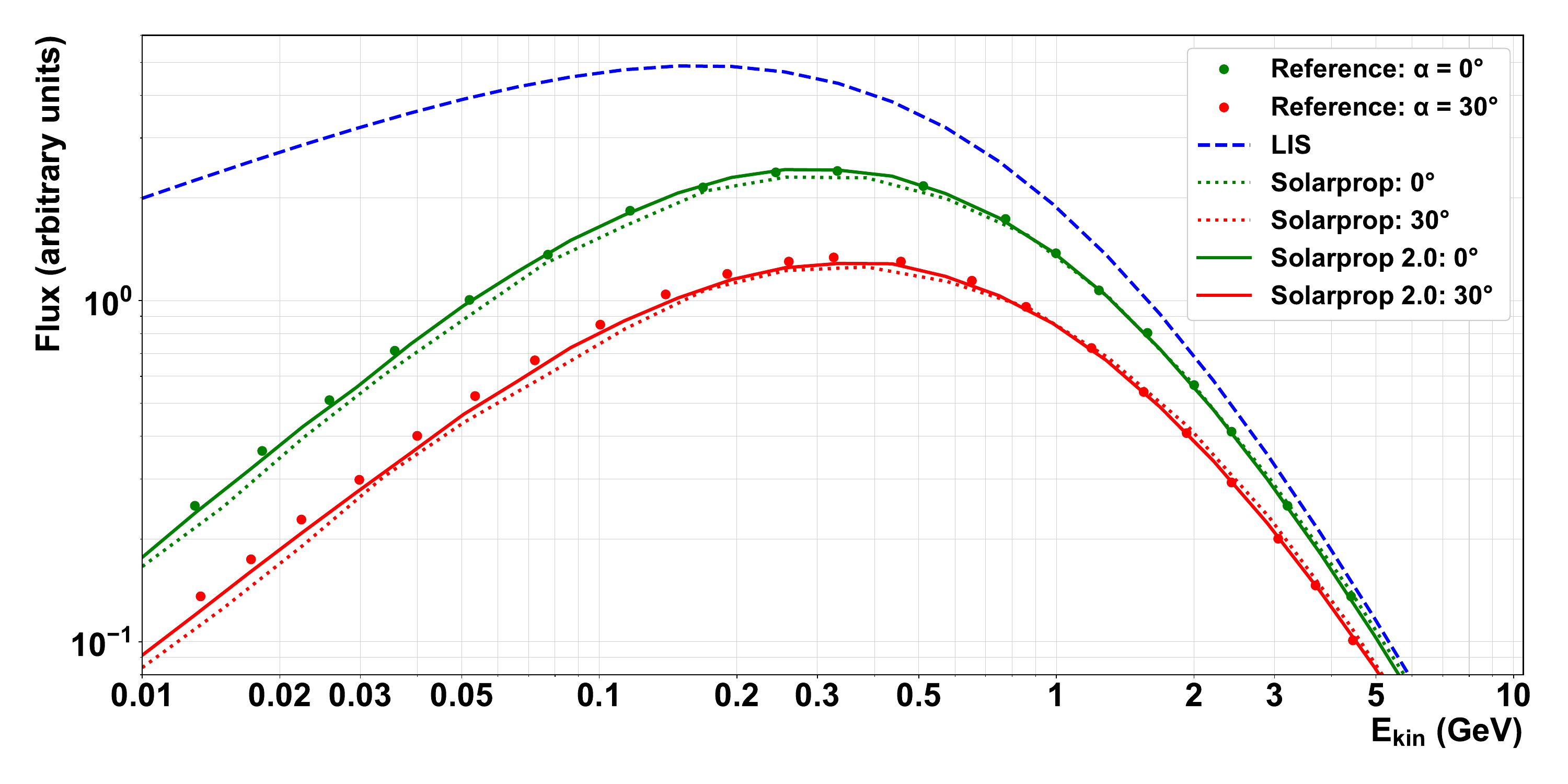}
  \caption{Validation of \solarproptwo{} with the model of Burger and
    Potgieter~\cite{Burger1989}. The points labeled ``Reference''
    have been digitized from the calculation results displayed in
    Figure~5 of Ref.~\cite{Burger1989}, showing results for $A<0$, for
    a flat neutral sheet and for a sheet with a tilt angle of
    $\alpha=\qty{30}{\degree}$.}
  \label{fig:bp}
\end{figure}
The fourth model aims at reproducing the results shown in Figure~5 of
a publication by Burger and Potgieter~\cite{Burger1989}. These results
were obtained with the same diffusion parameters as used in a
calculation by K\'ota and Jokipii~\cite{Kota1983}, but with the new
implementation of current sheet drift by Burger and Potgieter, see
also Ref.~\cite{Burger1995}. The results of the calculation with
\solarproptwo{} are in good agreement with the reference and the
agreement is improved with respect to vanilla \solarprop{}
(Fig.~\ref{fig:bp}). All Python scripts used to generate the figures
shown here for validation are found in the code repository.

\subsection{Three-dimensional models with wavy current sheet}
\subsubsection{Model of Burger (2012)}
\label{sec:burger2012}
To test the set of three-dimensional SDEs, eq.~(\ref{eq:sdes3d}), and
the ability of describing current sheet drift in detail also for a
wavy current sheet, the model of Burger~\cite{Burger2012} was
implemented in \solarproptwo{}. The implementation will be called
\texttt{Burger2012} model in the following. The model uses a simple
Parker spiral for the heliospheric magnetic field,
eq.~(\ref{eq:parkerspiral}). The diffusive part of the model is
straightforward, with the diffusion coefficients parallel and
perpendicular to the magnetic field given by
\begin{equation}
  \label{eq:burger2012:kappa}
  \kappa_\parallel=\kappa_0\,\beta{}\sqrt{R/(\qty{1}{GV}/c)}\,\frac{B_\oplus}{B}\quad\text{and}\quad\kappa_\perp=\num{0.05}\kappa_\parallel,
\end{equation}
respectively, with
$\kappa_0=\qty{5e21}{\centi\meter\squared\per\second}$. The
transformation from field-aligned to heliocentric coordinates is
achieved by the following equations~\cite{Burger2008}:
\begin{equation}
  \label{eq:trafo3d}
  \begin{aligned}
    \kappa_{rr}&=\kappa_\parallel\cos^2\psi+\kappa_{\perp}\sin^2\psi,\\
    \kappa_{r\phi}&=\kappa_{\phi{}r}=(\kappa_{\perp}-\kappa_\parallel)\cos\psi\sin\psi,\\
    \kappa_{\theta\theta}&=\kappa_{\perp},\\
    \kappa_{\phi\phi}&=\kappa_\parallel\sin^2\psi+\kappa_{\perp}\cos^2\psi,\\
  \end{aligned}
\end{equation}
where the spiral angle $\psi$ is defined by $\tan\psi=\Gamma$.

The model focuses on a detailed treatment of the drift along the
wavy heliospheric current sheet. Gradient and curvature drifts are
reduced in the vicinity of the current sheet by a prefactor
$\tanh(k(\theta^\prime-\theta)\cos\nu)$, where $\nu$ is the angle
between the normal to the current sheet and the $-\vec{e}_\theta$
direction, $k$ is a rigidity-dependent scale, and $\theta^\prime$
describes the location of the current sheet, in this model taken to be
\begin{equation}
  \label{eq:burger2012:thetaprime}
  \theta^\prime=\frac{\pi}{2}-\arctan\left[\tan\alpha\sin\left(\phi+\phi_0+\frac{\Omega{}r}{V_{sw}}\right)\right].
\end{equation}
The relevant velocity
field describing the current sheet drift is then given by
\begin{equation}
  \label{eq:burger2012:hcsdrift}
  \vec{v}_\mathrm{hcs}=\frac{k}{r\cosh^2(k(\theta^\prime-\theta)\cos\nu)}\kappa_A\,\vec{e}_{ns},\quad\text{with}\,\,\kappa_A=\frac{pv}{3qB},
\end{equation}
where the unit vector in the direction of the current sheet drift is
\begin{equation}
  \label{eq:burger2012:hcse}
  \vec{e}_{ns}=\sin\psi\cos\nu\,\vec{e}_{r}+\sin\nu\,\vec{e}_{\theta}+\cos\psi\cos\nu\,\vec{e}_{\phi}.
\end{equation}
\begin{figure}[tpb]
  \centering
  \includegraphics[width=\textwidth]{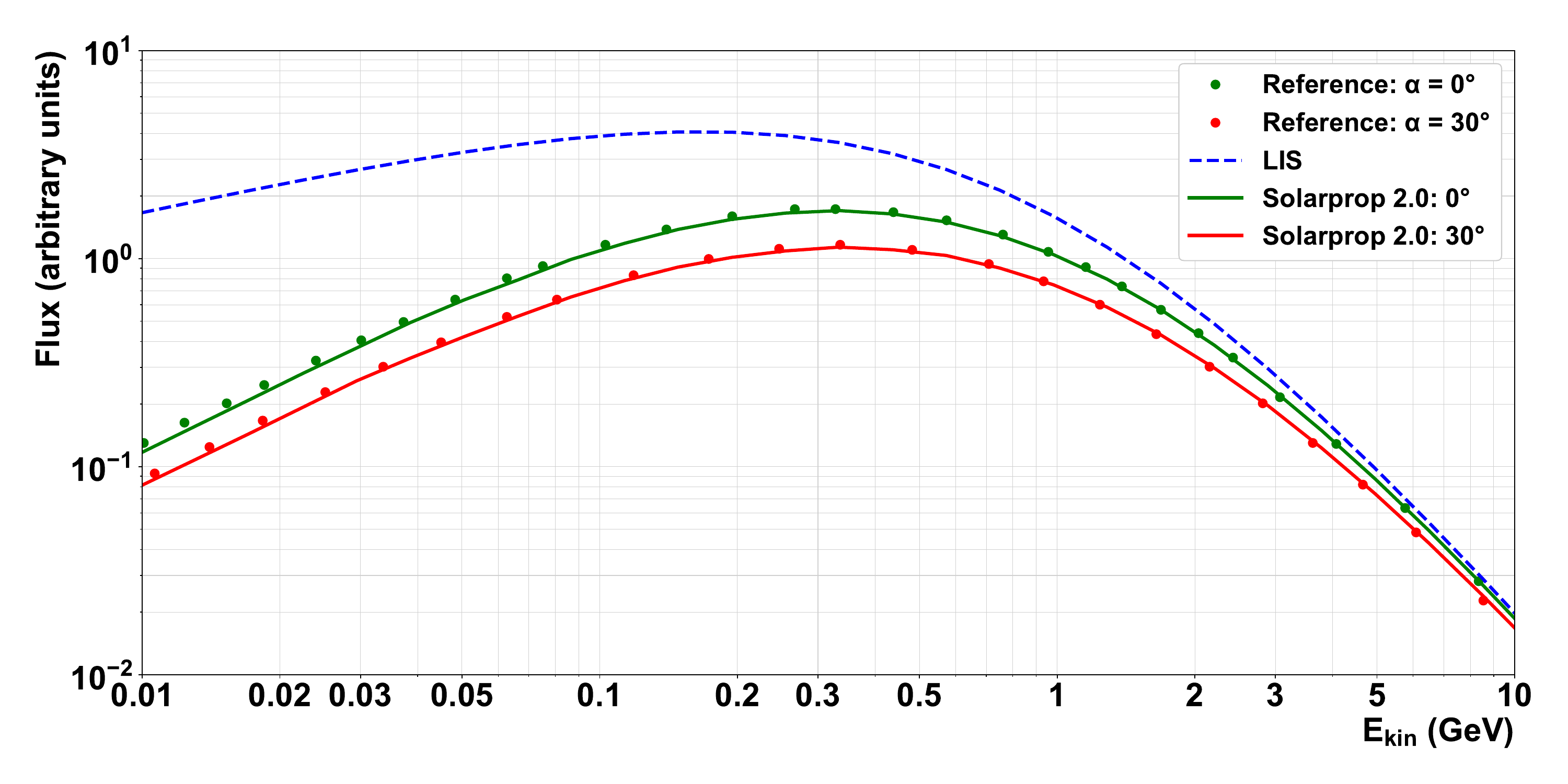}
  \caption{Validation of \solarproptwo{} with the model of
    Burger~\cite{Burger2012}. The points labeled ``Reference'' have
    been digitized from the calculation results displayed in Figure~2
    of Ref.~\cite{Burger2012}, showing results for protons at $A<0$,
    for a flat neutral sheet and for a current sheet tilt angle of
    $\alpha=\qty{30}{\degree}$.}
  \label{fig:burger2012:flux}
\end{figure}

\begin{figure}[tpb]
  \centering
  \includegraphics[width=\textwidth]{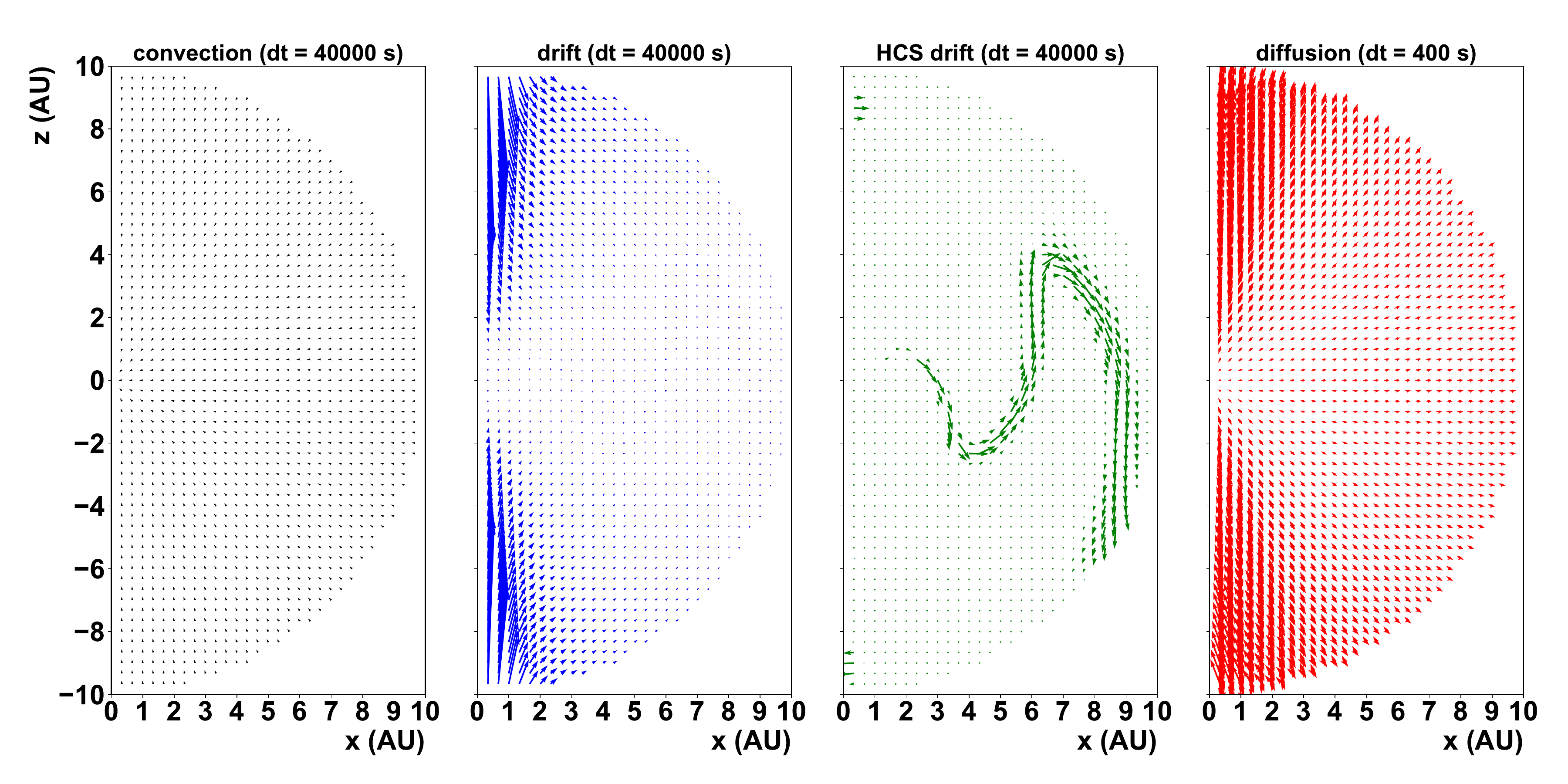}
  \caption{Illustration of different transport processes in the
    \texttt{Burger2012} model, for protons with a kinetic energy of
    \qty{100}{MeV}, $A<0$ polarity of the heliospheric magnetic field,
    and a current sheet tilt angle of $\alpha=\qty{30}{\degree}$. From
    left to right, the panels show the contributions of convection,
    gradient and curvature drifts, drift along the heliospheric
    current sheet, and diffusion, respectively. The corresponding time
    steps are indicated above each panel. Note that the vectors are
    shown for backward-in-time propagation. For diffusion, the arrows
    show the endpoints for ten different sets of random numbers, to
    illustrate the fluctuations inherent in the random nature of the
    process.}
\label{fig:burger2012:arrows}
\end{figure}

\begin{figure}[tpb]
  \centering
  \includegraphics[width=0.31\textwidth]{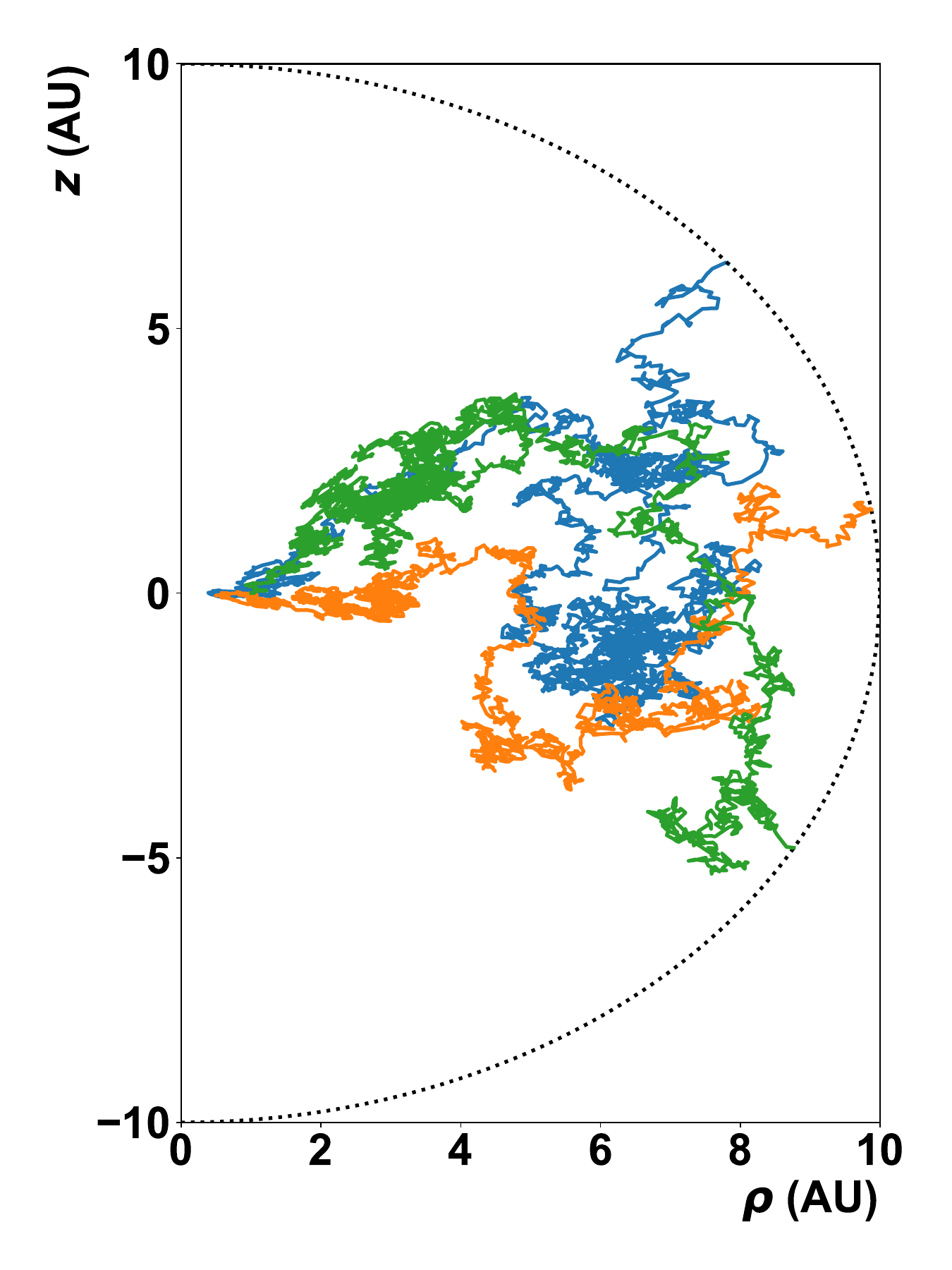}
  \includegraphics[width=0.31\textwidth]{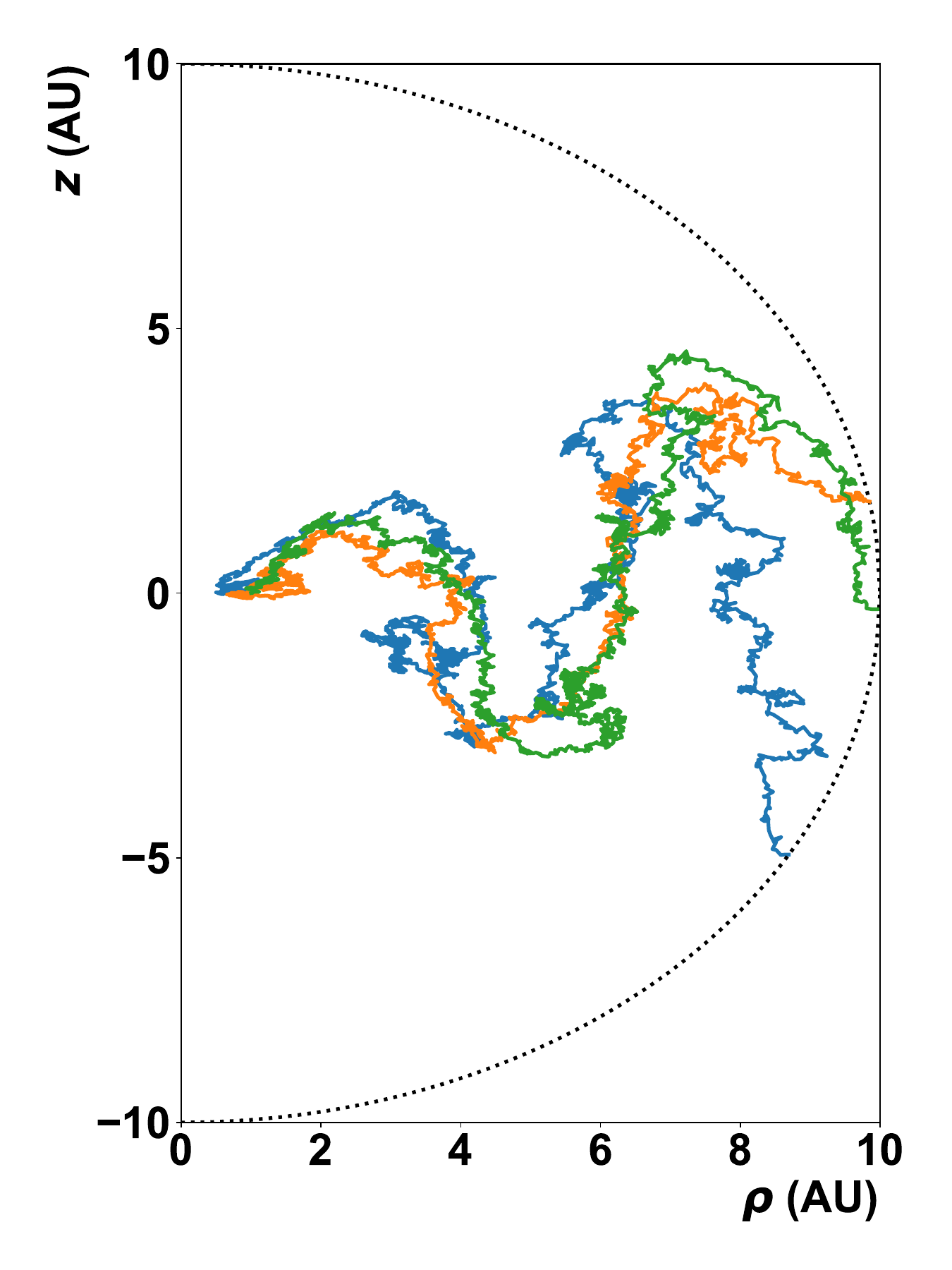}
  \includegraphics[width=0.31\textwidth]{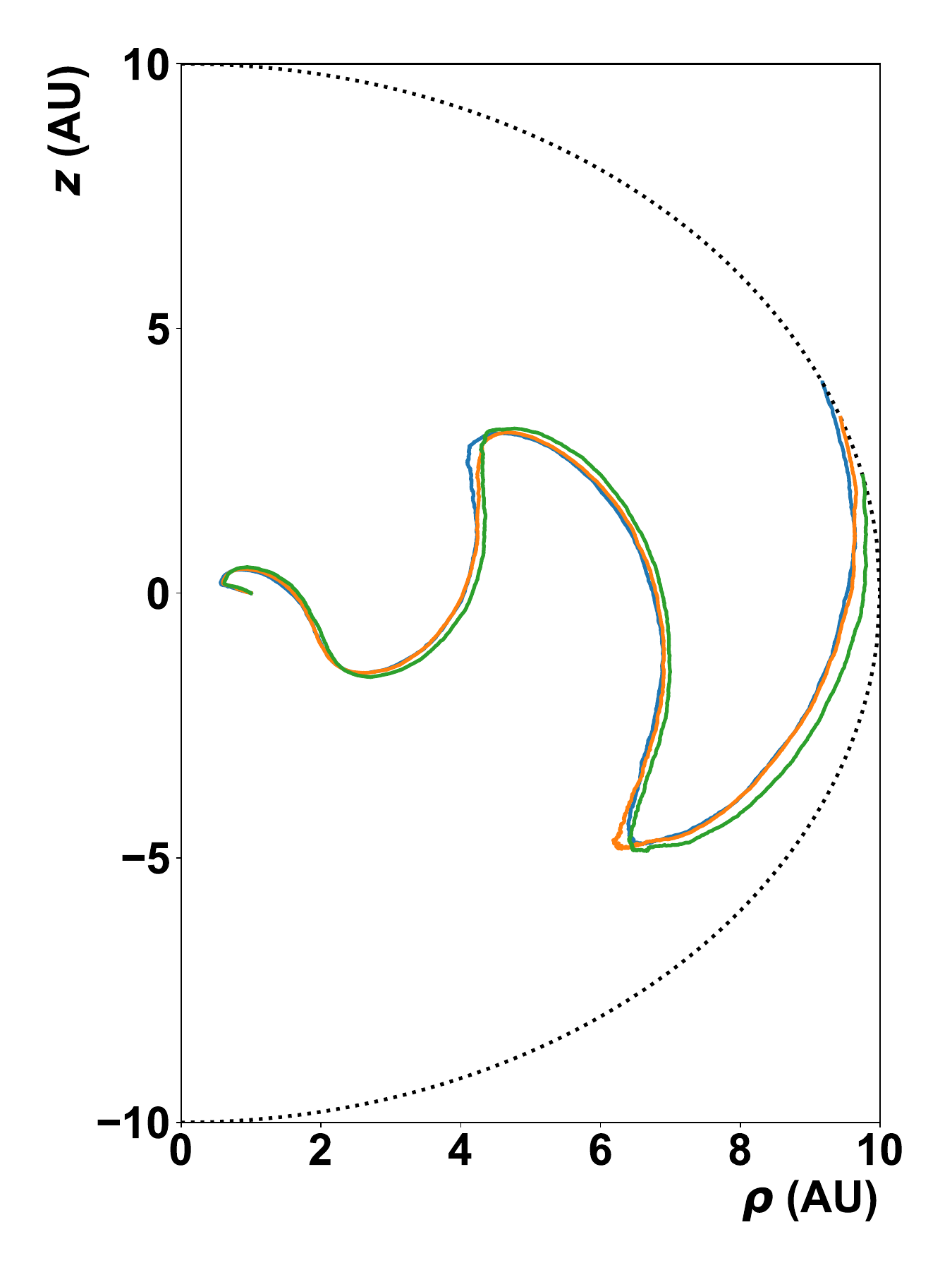}
  \caption{Illustration of pseudo-particle trajectories in the
    \texttt{Burger2012} model, for protons with a kinetic energy of
    \qty{100}{MeV}, $A<0$ polarity of the heliospheric magnetic field,
    and a current sheet tilt angle of $\alpha=\qty{30}{\degree}$. Each
    panel shows three realizations of pseudo-particle trajectories in
    different colors. From left to right, the diffusion scale factors
    are $\kappa_0$, $\num{0.3}\kappa_0$, and $\num{0.001}\kappa_0$,
    respectively. The horizontal axes show $\rho=\sqrt{x^2+y^2}$.}
\label{fig:burger2012:trajectories}
\end{figure}

For the full set of the relevant equations, the reader is kindly
referred to Ref.~\cite{Burger2012}. The results of the
calculation\footnote{Note that the LIS quoted in eq.~(20) of
  Ref.~\cite{Burger2012} does not match the curve shown in their
  Fig.~2. Instead, the spectral shape of the LIS used by K\'ota and
  Jokipii~\cite{Kota1983} is consistent with the Figure and has been
  taken for the validation study.} with \solarproptwo{} are in
excellent agreement with the reference
(Fig.~\ref{fig:burger2012:flux}).\newline
\par
This particular model presents a good opportunity to showcase the
visualization features of \solarproptwo{}: To understand and check the
contributions of the individual transport processes described by
Parker's equation, the \texttt{solarprop\_arrows} tool can be used to
plot the corresponding vector fields for a regular grid inside the
heliosphere, such as the ones shown in
Figure~\ref{fig:burger2012:arrows}. The arrows indicate the start and
end point of a pseudo-particle step, or several random steps for the
case of diffusion, for a given time increment. In this example, the
panel for gradient and curvature drifts shows the dampening effect of
the $\tanh$ prefactor in the vicinity of the wavy current sheet, while
the changing directionality and amplitude of the current sheet drift
field is clearly seen in the corresponding panel. The tool
\texttt{solarprop\_trajectories} produces visualizations of individual
pseudo-particle trajectories. In
Figure~\ref{fig:burger2012:trajectories}, three different random
realizations of trajectories are shown, for diminishing values of the
scale factor for the diffusion coefficient. It is seen that for the
nominal value of the diffusion coefficient, the particle transport is
dominated by diffusion, while drifts effectively confine the particles
to the angular regions covered by the current sheet, for the case of
$qA<0$ shown here. When the diffusion coefficient is set to an
artificially low value, diffusion can no longer contribute in a
significant way, and particle drift along the current sheet becomes
the only relevant process.

\subsubsection{Model of Strauss et al. (2012)}
An arguably even more detailed approach to modeling drifts along the
heliospheric current sheet was adopted by Strauss et
al.~\cite{Strauss2012}. In their prescription, the latitudinal extent
of the current sheet is taken to be
\begin{equation}
  \label{eq:strauss:hcs}
  \theta^\prime=\frac{\pi}{2}+\arcsin\left[\sin\alpha\sin\left(\phi-\phi_0+\frac{\Omega{}r}{V_{sw}}\right)\right].
\end{equation}
Before each step of a pseudo-particle during its propagation through
the heliosphere, the point on the wavy current sheet closest to its
position is located. Let $\vec{e}_r$, $\vec{e}_\theta$, and
$\vec{e}_\phi$ be the unit vectors in heliocentric spherical
coordinates in radial, polar, and azimuthal direction, respectively,
for the point of closest approach on the current sheet. Let $d$ be the
distance from the pseudo-particle to the point of closest approach on
the current sheet, and $v$ its velocity. The current sheet drift
velocity is then taken as
\begin{equation}
  \label{eq:strauss:vnsvector}
  \vec{v}_\mathrm{hcs}=qAv_\mathrm{hcs}\left(\cos(\pm\beta^\prime)\sin\psi\vec{e}_r+\sin(\pm\beta^\prime)\vec{e}_\theta+\cos(\pm\beta^\prime)\cos\psi\vec{e}_\phi\right),
\end{equation}
with a magnitude given by
\begin{equation}
  \label{eq:strauss:vns}
  v_\mathrm{hcs}=\left(\num{0.457}-\num{0.412}\frac{d}{r_L}+\num{0.0915}\frac{d^2}{r_L^2}\right)v,
\end{equation}
whenever the particle is within two Larmor radii $r_L=R/B$ of the
current sheet. The angle $\beta^\prime$ is the local inclination of
the current sheet with respect to the radial direction, as defined in
detail in Ref.~\cite{Strauss2012}, and $\psi$ is again the spiral
angle. At each location, the sign of $\beta^\prime$ determines whether the
particle drifts upward or downward along the current sheet.

A model with this treatment of the current sheet drift at its core was
implemented in \solarproptwo{}. The implementation will be referred to
as \texttt{Strauss2012} in the following. It should be pointed out
that locating the closest point to the particle on the wavy current
sheet is in fact an intricate procedure due to the complicated shape
of the current sheet. The implementation uses the Nelder-Mead simplex
algorithm provided by the GNU Scientific Library, with the current
particle position as the starting point. Care has been taken to find
good values for the initial step sizes.

\begin{figure}[tpb]
  \centering
  \includegraphics[width=0.75\textwidth]{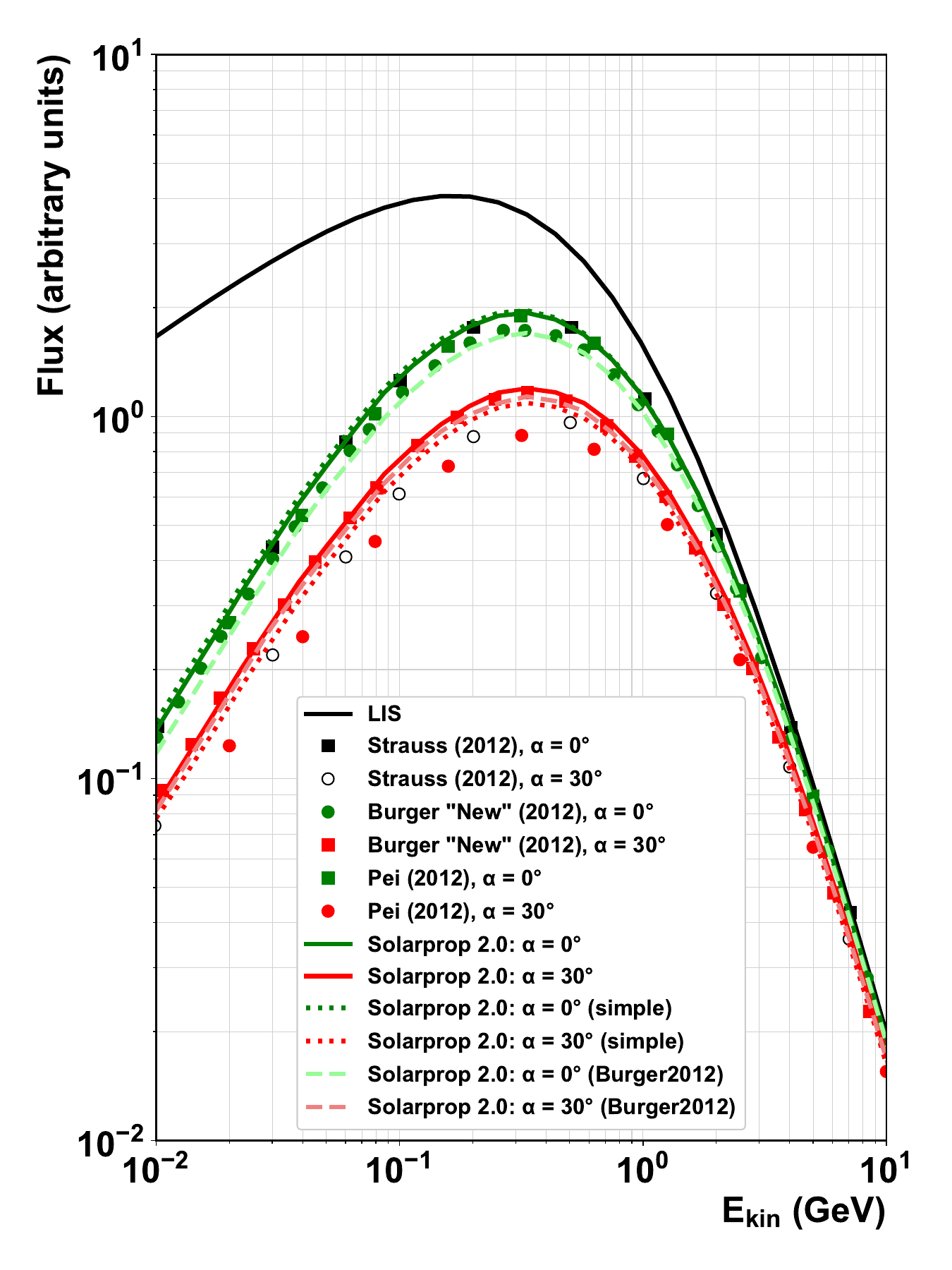}
  \caption{Comparison of \solarproptwo{} calculations to the results
    shown in Fig.~15 of Ref.~\cite{Strauss2012}, labeled here as
    ``Strauss (2012)''. Curves are shown for protons at $A<0$, for
    tilt angles of \qty{0}{\degree} and \qty{30}{\degree},
    respectively. Markers labeled ``Burger'' and ``Pei'' have been
    digitized from the relevant curves in Fig.~2 of
    Ref.~\cite{Burger2012} and Fig.~5 of Ref.~\cite{Pei2012},
    respectively. Please see the text for a discussion of the
    different curves from \solarproptwo{}.}
\label{fig:strauss:fig15}
\end{figure}

Ref.~\cite{Strauss2012} contains results for two different model
realizations, shown in their Figures~15 and~5, respectively, and these
will be investigated in turn. First, the results of \solarproptwo{}
calculations are compared to those shown in Fig.~15 of
Ref.~\cite{Strauss2012}, for protons and $A<0$ polarity. In this case,
the description of diffusion and the adopted model parameters are
identical to the \texttt{Burger2012} model implementation described in
section~\ref{sec:burger2012}. As seen in Fig.~\ref{fig:strauss:fig15},
a good match is obtained for the case of a flat current sheet,
$\alpha=\qty{0}{\degree}$. For $\alpha=\qty{30}{\degree}$ however,
\solarproptwo{} yields a flux somewhat higher than the reference
flux. To put this discrepancy into perspective, published results from
three different models by Strauss et al.~\cite{Strauss2012}, Pei et
al.~\cite{Pei2012}, and Burger~\cite{Burger2012}, all trying to
achieve a detailed treatment of drift along a wavy current sheet are
compared to each other in Fig.~\ref{fig:strauss:fig15}: There is
substantial spread in the predicted level of modulation \footnote{Note
  that the curves labeled ``Burger'' in Fig.~15 of
  Ref.~\cite{Strauss2012} are lower than the corresponding curves in
  Fig.~2 of the original publication, Ref.~\cite{Burger2012}.}  for
the case of a tilted sheet, $\alpha=\qty{30}{\degree}$.

To validate the rather involved treatment of the current sheet drift
further, two different alternative approaches have been adopted for
the calculation of the current sheet drift velocity
$\vec{v}_\mathrm{hcs}$. In the first approach, referred to as the
\emph{simple} case in the following as well as in
Fig.~\ref{fig:strauss:fig15}, the current sheet drift speed in
eq.~(\ref{eq:strauss:vnsvector}) is replaced by an average speed
inside the region affected by current sheet drift, which is defined by
the tilt angle and the particle's Larmor radius. Specifically, for
$\alpha>0$, the recipe given in section 4.2 of Ref.~\cite{Burger1995}
is followed:
\begin{equation}
  \label{eq:simpledrift}
  \vec{v}_\mathrm{hcs}=qA\,\frac{v}{6}\,\cos\alpha\,\frac{\Delta\theta}{\sin(\alpha+\Delta\theta)}\vec{e}_r,\quad\text{with}\quad
  \Delta\theta=\frac{2|R|V_{sw}}{B_0r_0^2\Omega\cos\alpha},
\end{equation}
if $|\theta-\pi/2|<\alpha+\Delta\theta$, and $\vec{v}_\mathrm{hcs}=0$
outside this cone. For the flat case, $\alpha=0$, the even simpler
formula
\begin{equation}
  \label{eq:simpledriftflat}
  \vec{v}_\mathrm{hcs}=
  \begin{cases}
    qA\,\frac{v}{6}\,\vec{e}_r & d\leq{}2r_L,\\
    0 & \text{else},
  \end{cases}
\end{equation}
is used, where $d=r\cos\theta$. In the second approach, the results
for the \texttt{Burger2012} model implementation
(cf.~Fig.~\ref{fig:burger2012:flux}) are used. It turns out that the
spread between these approaches is much smaller than the one between
the three published models used here for comparison. This is a strong
indication that the implementation of the detailed current sheet drift
in \solarproptwo{} works reliably.

It is interesting to note that in the flat case, the
\texttt{Burger2012} model gives somewhat stronger modulation than the
\texttt{Strauss2012} model. This is to be expected since in the
latter, the current sheet drift speed is simply added to the gradient
and curvature drifts, while the former introduces the prefactors which
reduce the drift and lead to larger propagation times and
correspondingly increased energy loss, and the effect is largest for
the flat current sheet, when drifts are most effective.\newline
\par

\begin{figure}[tpb]
  \centering
  \includegraphics[width=0.95\textwidth]{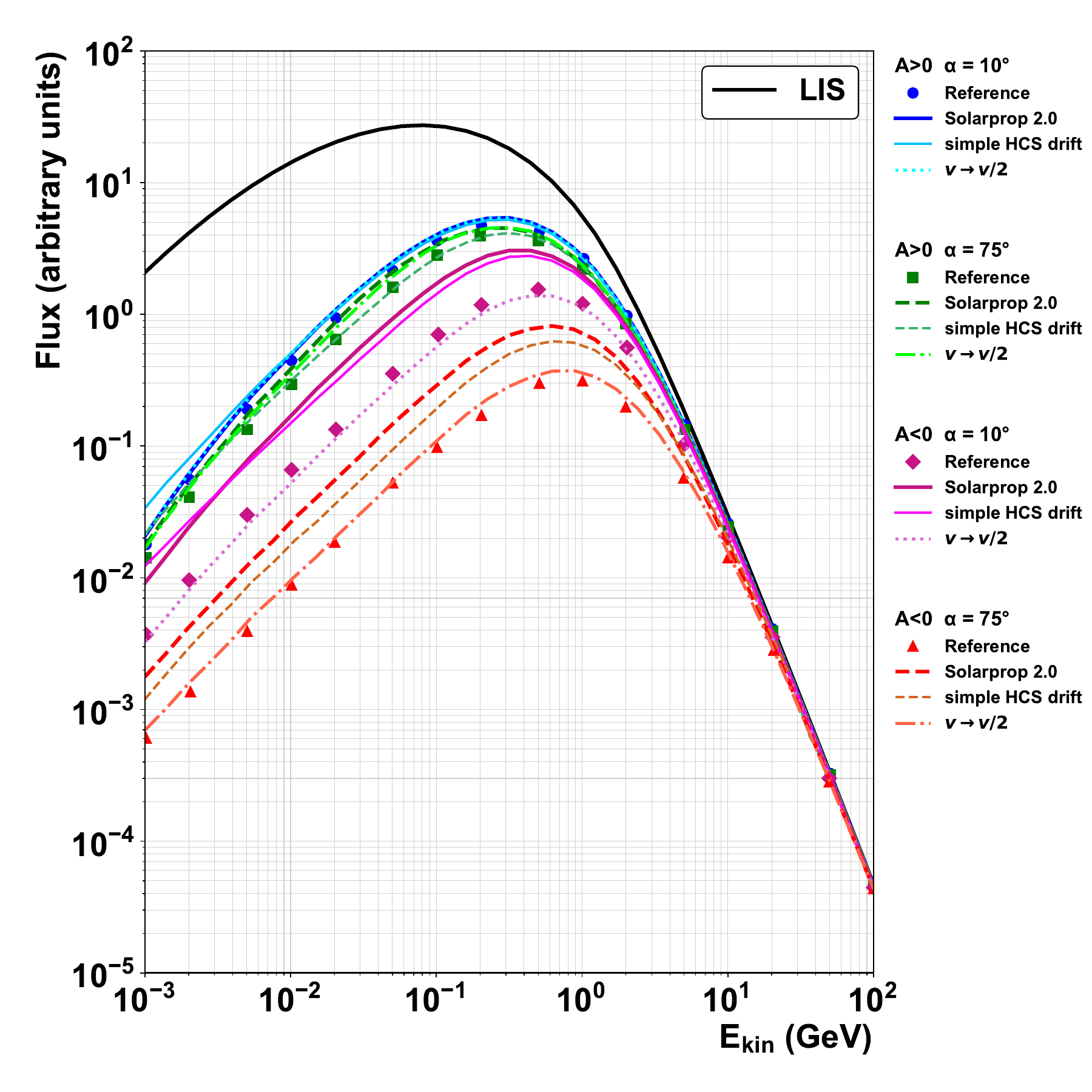}
  \caption{Comparison of \solarproptwo{} calculations to the reference
    results, digitized from Fig.~5 of Ref.~\cite{Strauss2012}, for
    four different combinations of magnetic field polarity $A$ and
    current sheet tilt angle $\alpha$. The thin curves labeled
    ``simple HCS drift'' use the simplified treatment according to
    eq.~(\ref{eq:simpledrift}) for the current sheet drift, while for
    the curves labeled ``$v\to{}v/2$'', the drift velocity in
    eq.~(\ref{eq:strauss:vns}) has been reduced by half.}
\label{fig:strauss:fig5}
\end{figure}
Second, the results of \solarproptwo{} calculations are compared to
the results shown in Fig.~5 of Ref.~\cite{Strauss2012}, obtained for a
heliosphere with a radius of \qty{100}{AU} and a constant solar wind
speed of $\vec{V}_{sw}=\qty{400}{\kilo\meter\per\second}\,\vec{e}_r$,
for protons assuming a parallel mean free path of\footnote{Note that
  $B$ and $B_0$ appear to be interchanged by accident in eq.~(21) of
  Ref.~\cite{Strauss2012}.}
\begin{equation}
  \label{eq:strauss:lambda}
  \lambda_\parallel=\lambda_0\left(\frac{B_0}{B}\right)\,\left(\frac{P}{P_0}\right),
\end{equation}
with $\lambda_0=\qty{0.05}{AU}$ and $P_0=\qty{1}{GV}/c$. The mean free
path is related to the parallel diffusion coefficient by
$\kappa_\parallel=\lambda_\parallel{}v/3$. Isotropic perpendicular
diffusion with $\lambda_{\perp{}r,\theta}=\num{0.02}\lambda_\parallel$
is included, like in the reference calculation. The parameterization
of the local interstellar spectrum (LIS) is given in
Ref.~\cite{Langner2004}. As seen in Fig.~\ref{fig:strauss:fig5}, the
calculation with \solarproptwo{} yields fluxes that are considerably
larger than those shown in Fig.~5 of Ref.~\cite{Strauss2012} for the
case of $A<0$, when protons enter the heliosphere predominantly
through the equatorial regions, where they encounter the current
sheet. As above, an alternative calculation with the current sheet
drift replaced by the simple case (eq.~(\ref{eq:simpledrift})) serves
to demonstrate that possible errors in the implementation of current
sheet drift cannot explain the magnitude of the observed
discrepancy. To illustrate the scale of the discrepancy, the absolute
value of the current sheet drift velocity in
eq.~(\ref{eq:strauss:vns}) was reduced by half, and this leads to
fluxes in agreement with those presented in Ref.~\cite{Strauss2012}.

\begin{figure}[tpb]
  \centering
  \includegraphics[width=0.75\textwidth]{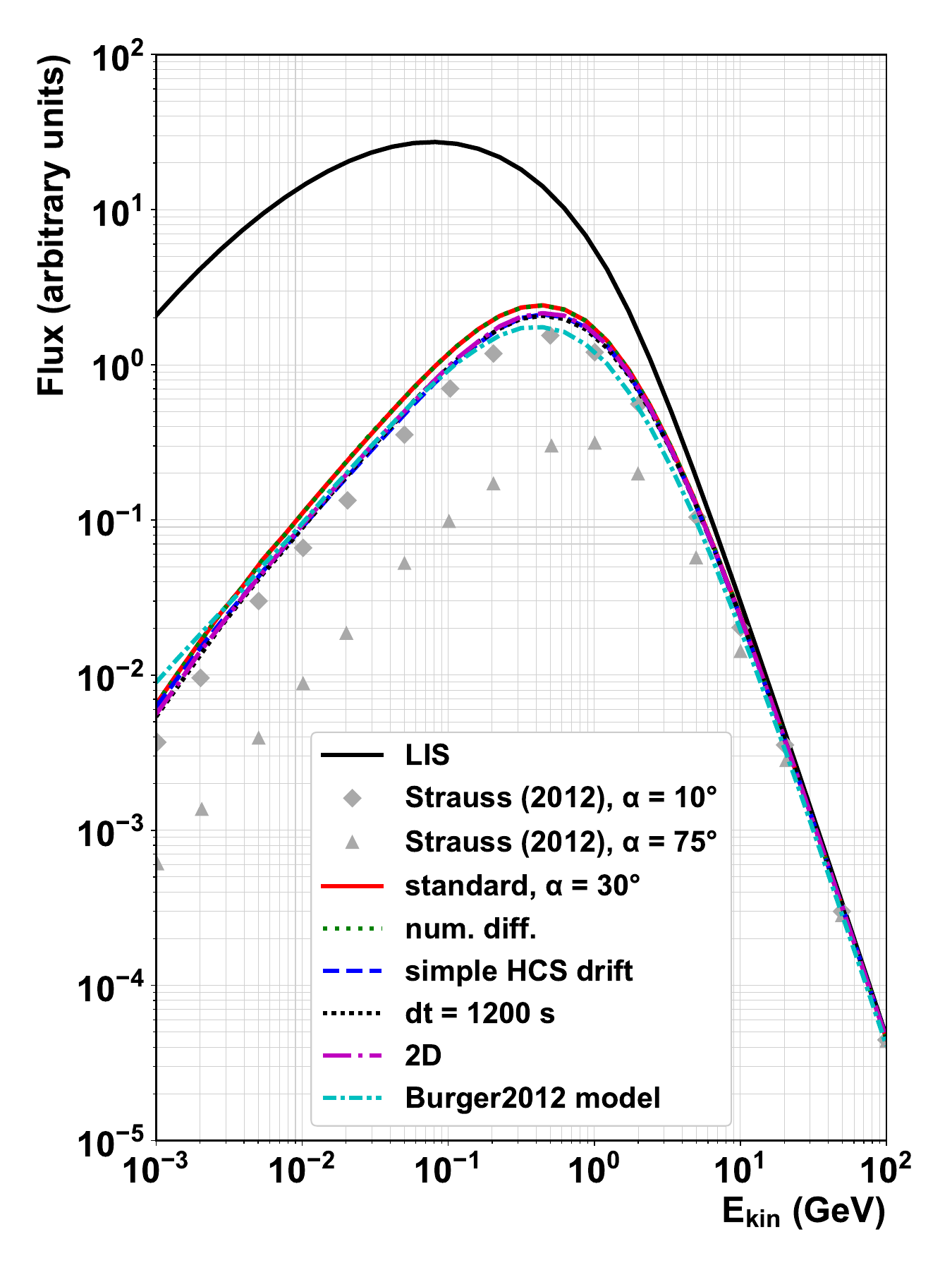}
  \caption{Results of the \solarproptwo{} implementation of the
    \texttt{Strauss2012} model, for protons, $A<0$, and a heliosphere
    extending to \qty{30}{AU}. The standard case with nominal settings
    and values is compared to several test cases with modifications
    testing the stability and consistency of the code, as discussed in
    the main text. Grey markers depict the same reference values for
    the $A<0$ case as in Fig.~\ref{fig:strauss:fig5}, to illustrate
    the amplitude of the predicted dependence on the tilt angle.}
\label{fig:strauss:variations}
\end{figure}
In the \solarproptwo{} implementation, the only meaningful change with
respect to the \texttt{Burger2012} model validated in the previous
section pertains to the treatment of the current sheet drift. To test
the stability and consistency of the implementation further, a study
was conducted where relevant aspects of the code were replaced one by
one with respect to the standard implementation. The same parameters
as for Fig.~\ref{fig:strauss:fig5} were used for this, except for the
extent of the heliosphere which was limited to \qty{30}{AU}, to avoid
instabilities in the \texttt{Burger2012} model. An intermediate tilt
angle of $\alpha=\qty{30}{\degree}$ has been chosen for the study.
The results for the flux calculated with the standard implementation
and with five different variations are shown in
Fig.~\ref{fig:strauss:variations}:
\begin{itemize}
\item ``\textit{num.~diff.}'': Numerical differentiation has been used
  to calculate the relevant derivatives of the diffusion coefficients
  in heliocentric spherical coordinates. Although this increases the
  run time, it reduces the amount of code needed to calculate the
  quantities in the SDEs considerably, thus eliminating potential bugs
  in the analytical calculation of the derivatives or their
  implementation in the source code. At the same time, this approach
  will automatically take care of implicit dependencies of any of the
  model parameters on the heliocentric coordinates $(r,\theta,\phi)$,
  which might be overlooked in manual calculations.

\item ``\textit{simple HCS drift}'': The detailed treatment of the current sheet
drift was replaced by the simplified one according to
eq.~(\ref{eq:simpledrift}).

\item ``\textit{dt = 1200 s}'': The integration time step was enlarged from the
standard value of $\mathrm{d}t=\qty{0.004}{days}=\qty{346}{s}$
to $\mathrm{d}t=\qty{1200}{s}$, to test the numerical stability of the
solution.

\item ``\textit{2D}'': The two-dimensional SDEs given in eq.~(\ref{eq:sdes2d}),
setting $\kappa_{r\theta}=0$, were used instead of the full
three-dimensional equations.

\item ``\textit{Burger 2012 model}'': The implementation presented in section
\ref{sec:burger2012} was used, but with the rigidity dependence and
overall scale of the diffusion coefficient as well as the ratio
$\kappa_\perp/\kappa_\parallel$ adjusted to match the settings of the
\texttt{Strauss2012} model.
\end{itemize}

While the results obtained with numerical differentiation are
identical to the standard case, the other replacements lead to small
changes in the calculated flux, as can be expected. But compared to
the scale of the observed discrepancy, the study demonstrates the
stability of the \solarproptwo{} implementation.

\begin{figure}[tpb]
  \centering
  \includegraphics[width=0.95\textwidth]{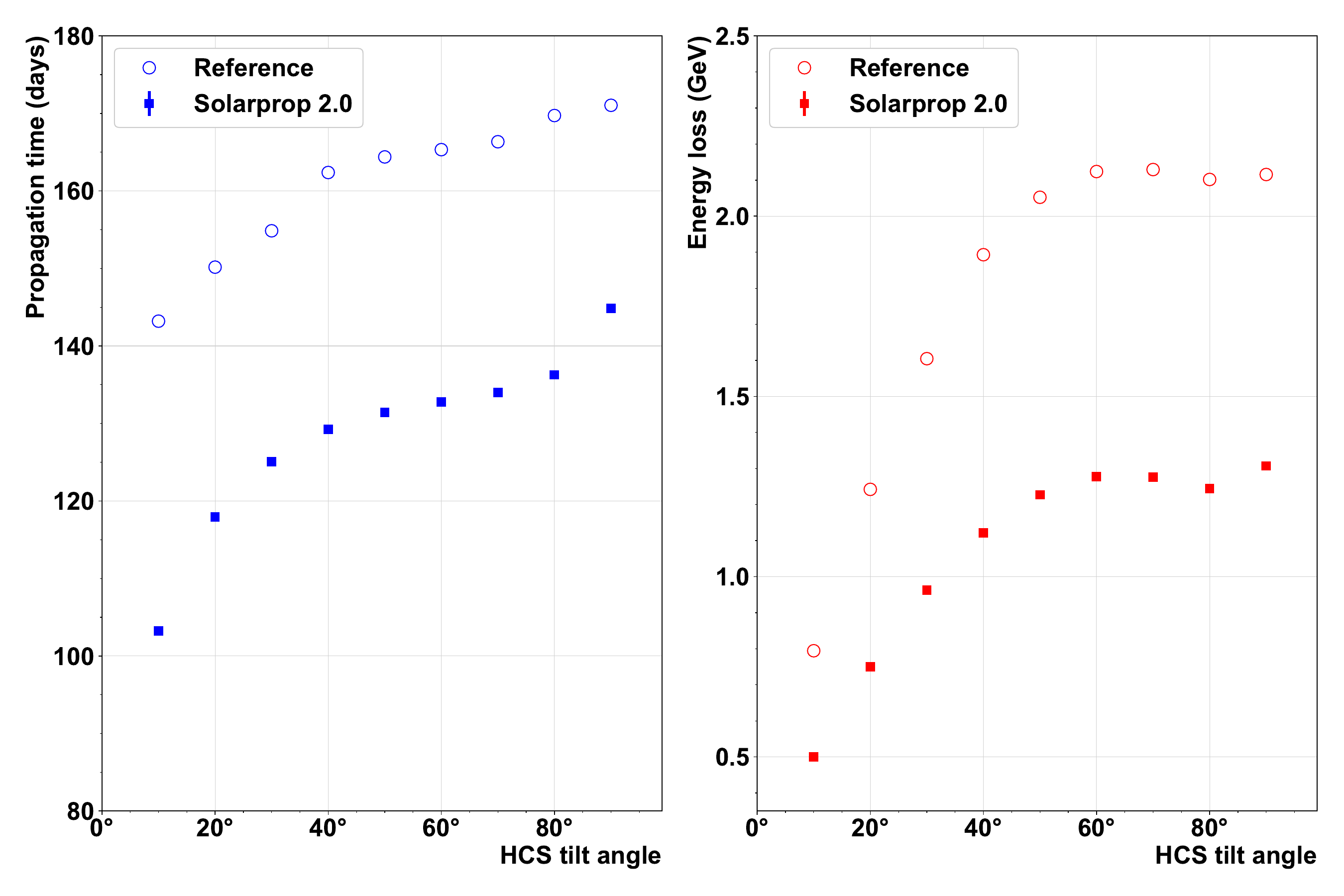}
  \caption{Average propagation time (\textit{left}) and energy loss
    (\textit{right}) of pseudo-particles in the \solarproptwo{}
    implementation of the \texttt{Straus2012} model, depending on the
    tilt angle, for protons with a kinetic energy of \qty{100}{MeV}
    and $A<0$. The uncertainties on the average values are smaller
    than the symbol sizes. The reference values have been digitized
    from Fig.~11 of Ref.~\cite{Strauss2012}.}
\label{fig:strauss:fig11}
\end{figure}
To investigate a possible root cause of the observed discrepancy, the
average propagation time and energy loss of pseudo-particles as a
function of the tilt angle have been calculated. The results are shown
in Fig.~\ref{fig:strauss:fig11}, compared to those given in Fig.~11 of
Ref.~\cite{Strauss2012}. Qualitatively, the same trends in the
dependence of the propagation times and energy losses on the current
sheet tilt angle are observed. But according to the reference
publication, the pseudo-particles propagate for much longer and
correspondingly lose more energy, which leads to considerably lower
fluxes. There is a possible explanation: When the point of closest
approach is wrongly located on the next-to-nearest fold of the current
sheet, the angle $\beta^\prime$ will receive the wrong sign and this effect
tends to send particles into loops along the current sheet (which are
of course not fully realized due to the diffusive component in the
propagation), thus prolonging the propagation time. This behavior was
sometimes observed when non-ideal settings for the initial step sizes
in the Nelder-Mead minimization of the distance to the current sheet
were employed.

The differences in the results between two implementations of the same
model revealed in this section highlight the necessity for a detailed
comparison. In the end, this can only be done at the level of the
source code.

\section{Model with modified magnetic field}
\label{sec:model-with-modified-mag-field}
It has been known for a long time that the simple Parker spiral
(eq.~(\ref{eq:parkerspiral})) would lead to large particle drifts in
the polar regions of the heliosphere at odds with experimental
observations. Two models from the line of development of the HelMod
modulation suite~\cite{Bobik2012,Boschini2018a} have been implemented
in \solarproptwo{}. They are centered around the modification of the
heliospheric magnetic field proposed by Jokipii and
K\'ota~\cite{Jokipii1989}, as also advocated by Heber et
al.~\cite{Heber1998}, and thus make use of the full set of SDEs for
the two-dimensional case (eqs.~(\ref{eq:sdes2d})). The two models
differ mainly in the spatial dependence of the diffusion coefficient
and its overall scale. The implementation of the first
model~\cite{Bobik2012} will be described in detail in the following as
an example. It is important to stress that the following is a
description of the implementation in \solarproptwo{}, which aims to be
as faithful as possible to the model in Ref.~\cite{Bobik2012}, except
that the backward-in-time approach has been used for the solution of
the SDEs for reasons of efficiency, and the data files for
time-dependent parameters have been downloaded from their respective
sources in their current state. Other known differences in the
implementation are mentioned explicitly in the following.


\begin{figure}[htp]
  \centering
  \includegraphics[width=0.55\textwidth]{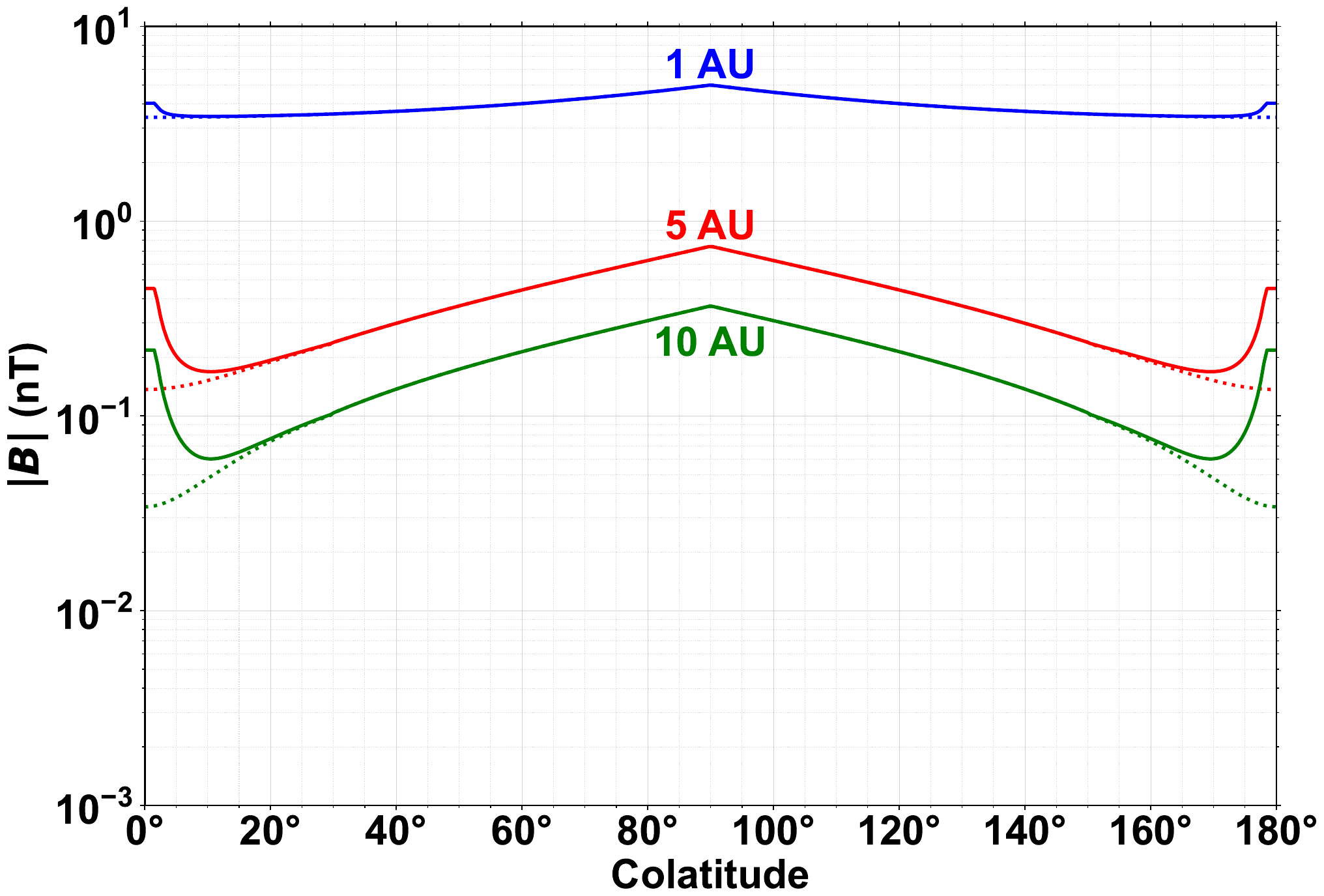}
  \includegraphics[width=0.2\textwidth]{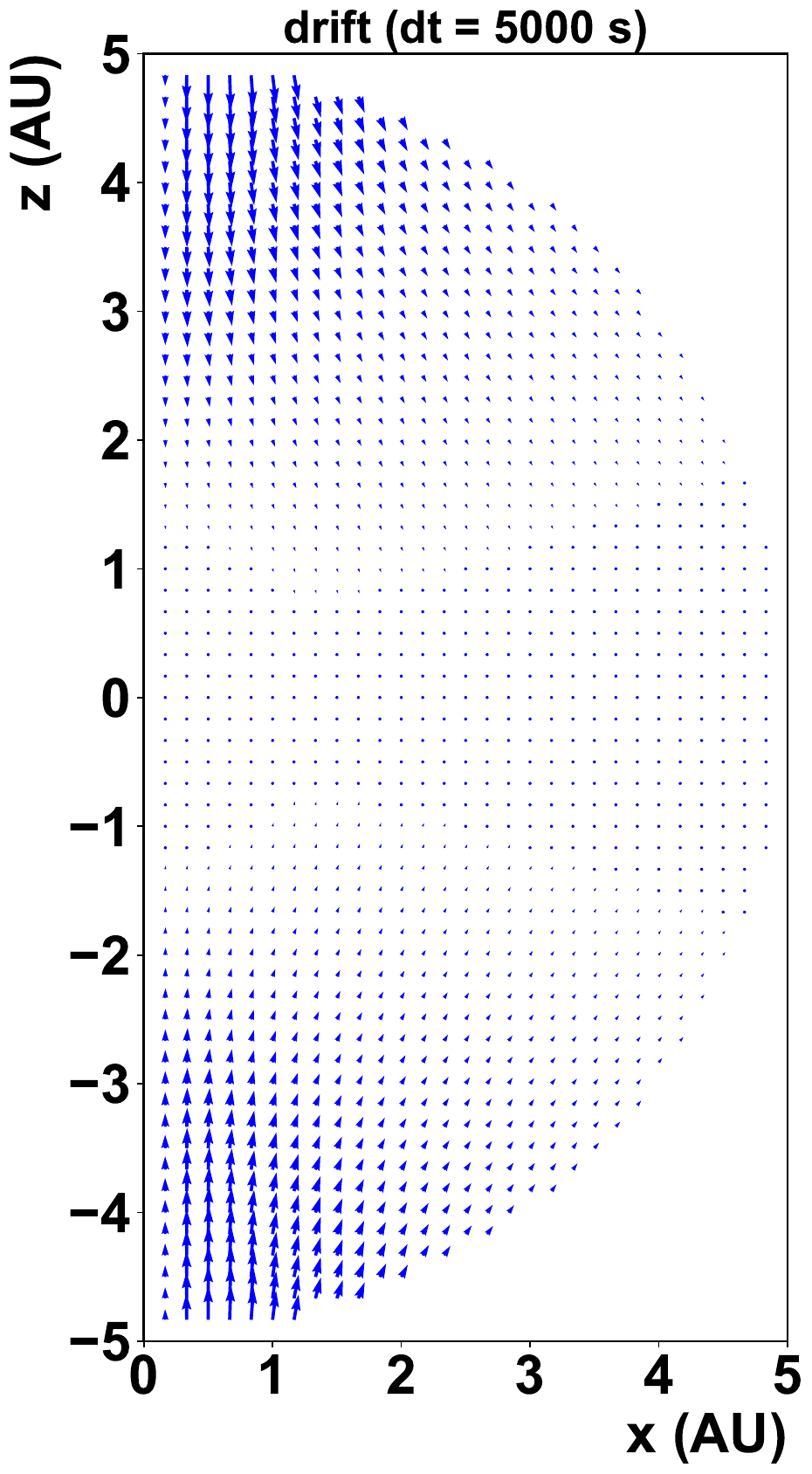}
  \includegraphics[width=0.2\textwidth]{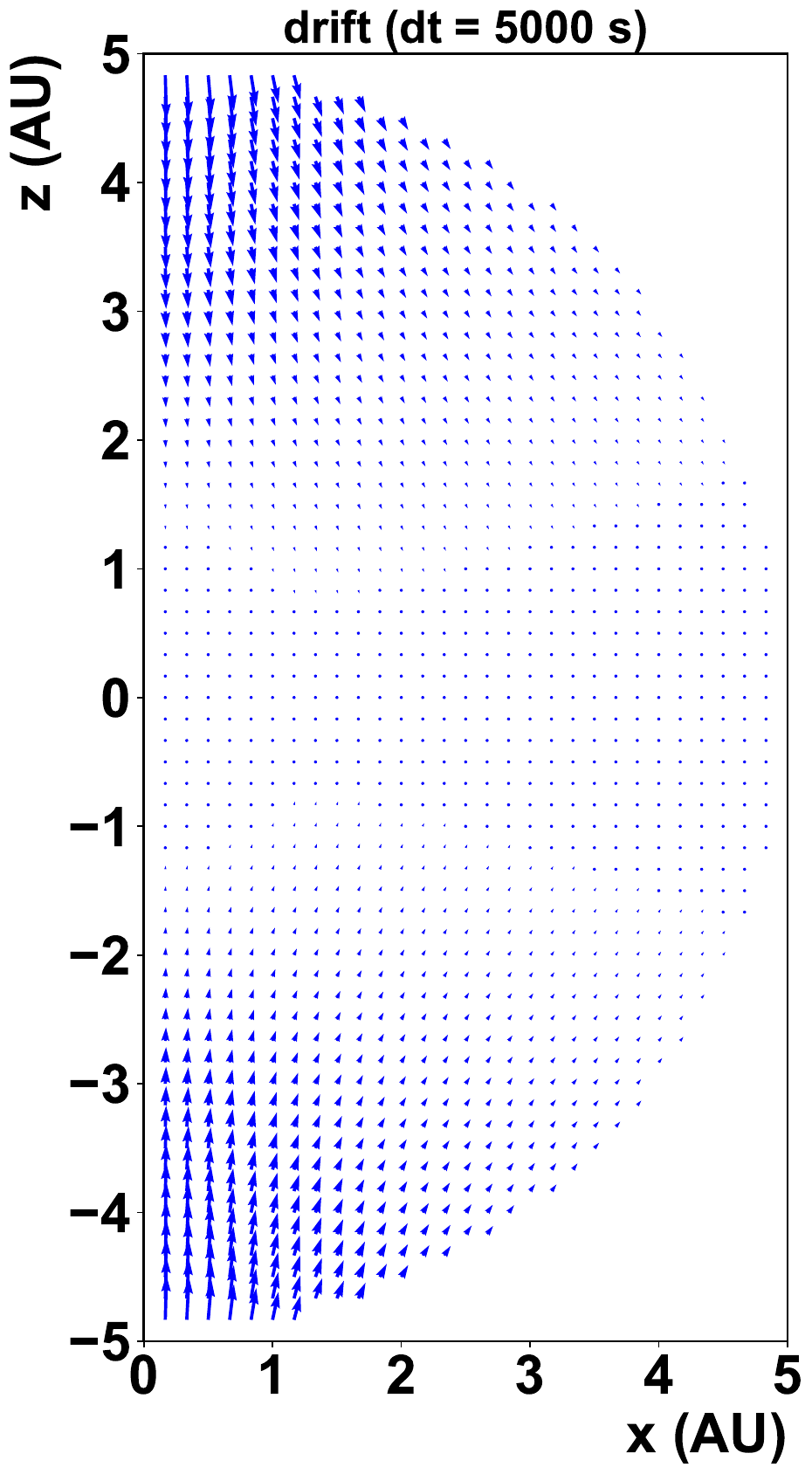}
  \caption{Left: Amplitude of the modified magnetic field according to
    eq.~(\ref{eq:magfield_jk}) vs colatitude $\theta$, at three
    different radial distances from the Sun, as labeled. The dotted
    lines show the respective amplitudes for the case
    $\delta(\theta)=0$, i.e. the usual Parker spiral. The other two
    plots visualize the effect of the magnetic field modification on
    the calculated $\mathrm{d}\vec{r}$ due to gradient and curvature
    drifts, for $A<0$ polarity, a time step of
    $\mathrm{d}t=\qty{5000}{s}$ and particles with a kinetic energy of
    \qty{0.5}{GeV}. The middle plot corresponds to the modified
    magnetic field, while the right plot is for
    $\delta(\theta)=0$. Note the differences in the polar regions.}
  \label{fig:helmod:magfield}
\end{figure}

The magnetic field is modified with respect to the Parker spiral in
the polar regions of the heliosphere and given by
\begin{equation}
  \label{eq:magfield_jk}
  \vec{B}=AB_0\frac{r_0^2}{r^2}\left[\,\vec{e}_r - \Gamma\,\vec{e}_\phi + \frac{\delta(\theta)}{r_\odot}r\,\vec{e}_\theta\right] \times \left[1-2H(\theta-\theta^\prime)\right],
\end{equation}
where the location of the heliospheric current sheet is determined by
eq.~(\ref{eq:location_hcs}). Taking the detailed specification
found in Ref.~\cite{Bobik2013} into account, the latitudinal
correction function for the polar regions is
\begin{equation}
  \label{eq:delta_jk}
  \delta(\theta)=\left[1-2H(\theta-\theta^\prime)\right]\times
  \begin{cases}
    0 & \qty{30}{\degree}\leq\theta\leq\qty{150}{\degree}, \\
    \delta_0/\sin\theta & \theta_m\leq\theta < \qty{30}{\degree}\,\,\text{or}\,\,
    \qty{150}{\degree}<\theta\leq\qty{180}{\degree}-\theta_m,\\
    \delta_0/\sin\theta_m & \text{else},
  \end{cases}
\end{equation}
with $\theta_m=\qty{1.7}{\degree}$ and the scale of the correction is
given by $\delta_0=\num{8.7e-5}$. Finally, $r_\odot=\qty{6.963e8}{m}$
denotes the solar radius, and the other quantities are defined as for
eq.~(\ref{eq:parkerspiral}). The amplitude of the modified magnetic
field is shown for three different values of $r$ in
Fig.~\ref{fig:helmod:magfield} (left) and compared to the simple
Parker spiral. Note that the curves for $r=\qty{5}{AU}$ and
$r=\qty{10}{AU}$ differ from the corresponding ones in Fig.~2 of
Ref.~\cite{Bobik2012}. To my understanding, the curves presented here
are correct. It is unclear if this discrepancy affects only the
visualization of the magnetic field amplitude or also the results of
the calculations given in Ref.~\cite{Bobik2012}. The other two plots
in Fig.~\ref{fig:helmod:magfield} illustrate the effect of the
magnetic field modification on drifts in the polar regions of the
heliosphere.

In periods of high solar activity, the velocity of the solar wind is
taken to be $V_{sw}=v_\mathrm{omni}$, where $v_\mathrm{omni}$ is the
27-day averaged value as extracted from NASA/GSFC's OMNI data set
through \texttt{OMNIWeb}~\cite{omniweb}
for the plasma flow speed. In all other periods, the solar wind speed
is
\begin{equation}
  \label{eq:sw_helmod}
  V_{sw}(\theta)=
  \begin{cases}
    \qty{760}{\kilo\meter\per\second} &
    \theta\leq\qty{30}{\degree}\,\,\text{or}\,\,\theta\geq\qty{150}{\degree}, \\
    v_\mathrm{omni}\,\cdot\,(1+|\cos\theta|) & \qty{30}{\degree}<\theta<\qty{150}{\degree}.\\
  \end{cases}
\end{equation}
Since no definition of high solar activity is given in
Ref.~\cite{Bobik2012}, a threshold of $N_\mathrm{sunspot}>150$ is used,
where $N_\mathrm{sunspot}$ is the 13-month smoothed monthly total
sunspot number from WDC-SILSO, Royal Observatory of
Belgium~\cite{SSN}. The model assumes a spherical heliosphere with a
radius of \qty{100}{AU}.

In field-aligned coordinates, the parallel component of the diffusion
tensor is given by
\begin{equation}
  \label{eq:helmod:kappaparallel}
  \kappa_\parallel=\beta\kappa_0K(R)\,\frac{B_\oplus}{3B}.
\end{equation}
In this equation, the overall scale factor $\kappa_0$ is a
time-dependent quantity whose value is calculated from an empirical
relation based on the smoothed sunspot number, as given in Eq.~(13)
and Table~1 of Ref.~\cite{Bobik2012}. There are four different
parameterizations depending on the heliospheric polarity $A$ and the
phase (ascending or descending activity), even though the exact time
ranges for these are not specified. The dependence of the
diffusion coefficient on the particle rigidity $R$ is given by
$K(R)=\mathrm{max}(1, R/(\qty{1}{GV}/c))$, and $B_\oplus$ is the
amplitude of the heliospheric magnetic field at Earth, as extracted
from the OMNI data set. The transverse diffusion coefficients are
given by $\kappa_{\perp{}r}=\rho\kappa_\parallel$ and
$\kappa_{\perp\theta}=\iota(\theta)\rho\kappa_\parallel$, with
$\rho=\num{0.05}$ and $\iota(\theta)=1$ in the equatorial region
($\qty{30}{\degree}\leq\theta\leq\qty{150}{\degree}$) and
$\iota(\theta)=10$ elsewhere.

The transformation from field-aligned to heliocentric coordinates
follows the prescription of Ref.~\cite{Burger2008} so that the
relevant components of the (symmetric) diffusion tensor are given by
\begin{equation}
  \label{eq:helmod:kappas}
  \begin{aligned}
    \kappa_{rr}&=\kappa_\parallel\left(\iota\rho\sin^2\xi+\cos^2\xi\,(\cos^2\psi+\rho\sin^2\psi)\right),\\
    \kappa_{\theta\theta}&=\kappa_\parallel\left(\iota\rho\cos^2\xi+\sin^2\xi\,(\cos^2\psi+\rho\sin^2\psi)\right),\\
    \kappa_{r\theta}&=\kappa_\parallel\left(\sin\xi\cos\xi\,(\cos^2\psi+\rho\sin^2\psi-\iota\rho)\right),\\
  \end{aligned}
\end{equation}
where the spiral angle $\psi$ is now defined by
$\tan\psi=-B_\phi/\sqrt{B_r^2+B_\theta^2}$, and the polar field angle
$\xi$ is defined by $\tan\xi=B_\theta/B_r$.

The drift velocity $\vec{V}_{dr}$ is given by
\begin{equation}
  \label{eq:helmod:drift}
  \vec{V}_{dr}=\tilde{f}(\theta)\,\vec{\nabla}\times\left(\kappa_A\frac{\vec{B}}{B}\right)+\left(\frac{\partial\tilde{f}(\theta)}{\partial\theta}\right)\frac{\kappa_A}{r}\,\vec{e}_\theta\times\left(\frac{\vec{B}}{B}\right),
\end{equation}
with $\kappa_A=Rv/3B$. The first term describes gradient and curvature
drifts, while the second term effectively accounts for drifts along the
heliospheric current sheet. The respective contributions are governed
by the transition function
\begin{equation}
  \label{eq:helmod:transitionf}
  \tilde{f}(\theta)=
  \begin{cases}
    (1/a_h)\arctan\left((1-2\theta/\pi)\tan{}a_h\right) & c_h<\pi/2, \\
    1-2H(\theta-\pi/2) & c_h=\pi/2, \\
  \end{cases}
\end{equation}
with $a_h=\arccos(\pi/(2c_h)-1)$ and
\begin{equation}
  \label{eq:helmod:ch}
  c_h=\frac{\pi}{2}-\frac{1}{2}\sin(\alpha+\Delta\theta_\mathrm{hcs}),
\end{equation}
where $\alpha$ is the tilt angle of the current sheet, and
$\Delta\theta_\mathrm{hcs}=2r_L/r$, where $r_L$ is again the Larmor
radius of the particle. The time-dependent value of the tilt angle is
extracted from the Wilcox Solar Observatory
data~\cite{tiltangle,Hoeksema1995}. For quiet conditions, $c_h$ tends
to $\pi/2$ and $\tilde{f}(\theta)$ becomes a step function that goes
from $1$ to $-1$ at the heliospheric equator. In the \solarproptwo{}
implementation, $c_h$ is explicitly limited to be above $\pi/4$, to
keep the expression for $a_h$ well-defined. For the case
$c_h\to\pi/4$, $\tilde{f}(\theta)$ tends to a straight line with
$\tilde{f}(0)=1$ and $\tilde{f}(\pi)=-1$ at the heliospheric north and
south poles, respectively.

To account for the propagation of changes in heliospheric conditions
with the solar wind, the heliosphere is divided into 15 equidistant
sectors. Each consecutive sector goes back one Carrington rotation
(\qty{27.28}{days}) in time, and the relevant model parameters are
extracted for the point in time associated with each sector
accordingly.

It should be noted that the polar component in the magnetic field
causes the derivatives of the elements of the diffusion tensor in
heliocentric coordinates to be considerably more involved than in the
case of the simple Parker spiral. To reduce the potential for mistakes
in the calculation and implementation of the corresponding
expressions, \texttt{sympy} has been used for the symbolic calculation
of all relevant derivatives as well as for the generation of the
corresponding C++ source code using the \texttt{sympy.ccode} function,
both for the terms involving the diffusion coefficients of
eq.~(\ref{eq:helmod:kappas}) and their derivatives in
eq.~(\ref{eq:sdes2d}), and for the calculation of the drift velocity
components according to eq.~(\ref{eq:helmod:drift}).

\begin{figure}[tpb]
  \centering
  \includegraphics[width=0.95\textwidth]{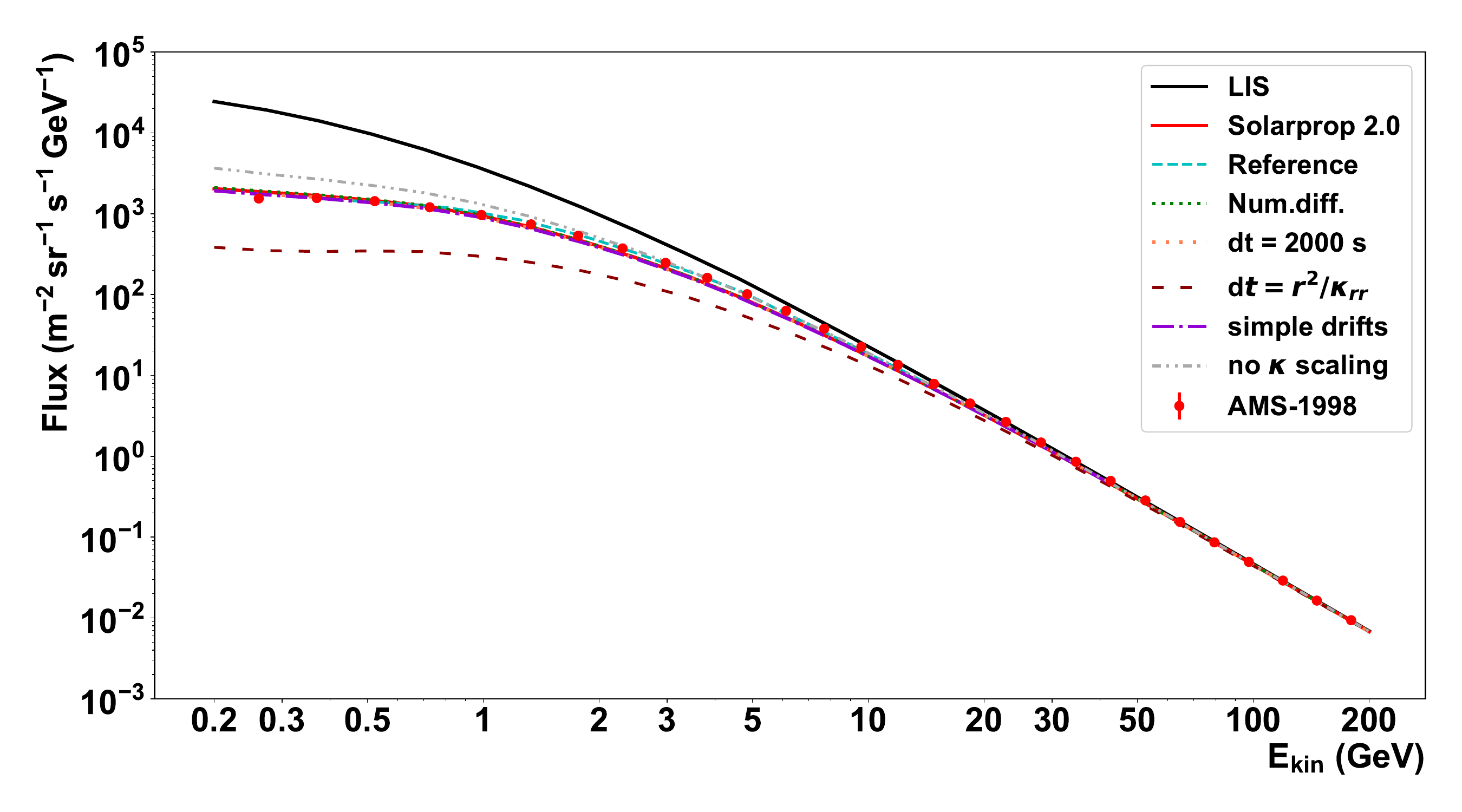}
  \caption{Comparison of \solarproptwo{} calculations to reference
    results digitized from Fig.~8 of Ref.~\cite{Bobik2012} and to
    AMS-01 data~\cite{Alcaraz2000a}. In addition, several variations
    of the model implementation are shown, as discussed in the
    text. Note that the diffusion coefficient $\kappa_0$ has been
    scaled by a factor \num{0.65} with respect to the reference
    value.}
  \label{fig:helmod:ams98}
\end{figure}

\begin{figure}[tpb]
  \centering
  \includegraphics[width=0.75\textwidth]{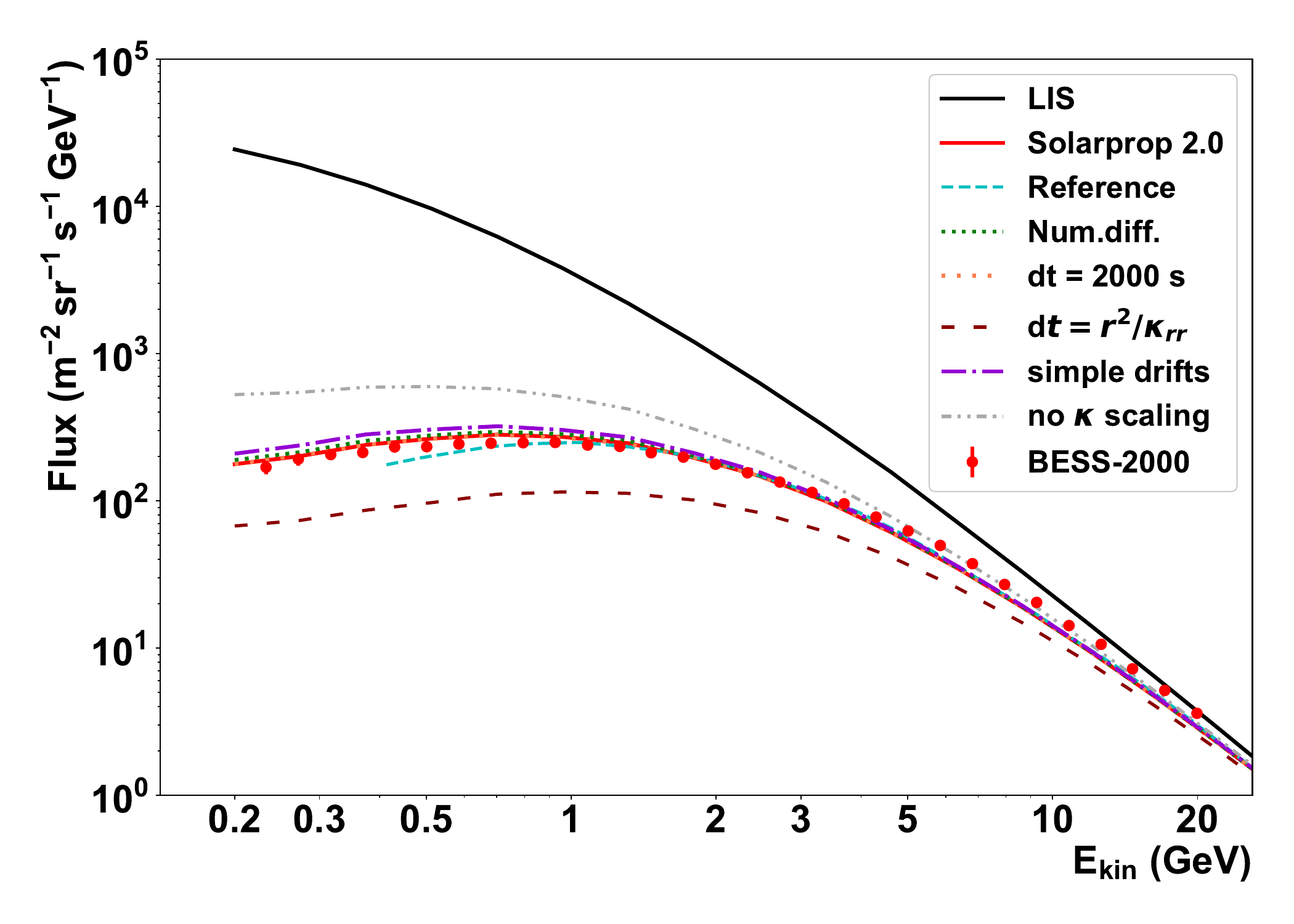}
  \caption{Comparison of \solarproptwo{} calculations to reference
    results digitized from Fig.~5 of Ref.~\cite{Bobik2012} and to BESS
    data~\cite{Shikaze2007}. In addition, several variations of the
    model implementation are shown, as discussed in the text. Note
    that the diffusion coefficient $\kappa_0$ has been scaled by a
    factor \num{0.65} with respect to the reference value.}
  \label{fig:helmod:bess00}
\end{figure}

To compare to the results of the reference
publication~\cite{Bobik2012}, calculations with the corresponding
\solarproptwo{} implementation have been performed for two points in
time as characteristic examples, first for the AMS-01 flight in
June~1998~\cite{Alcaraz2000a}, which took place during relatively
quiet heliospheric conditions with $A>0$ polarity, and second for the
balloon-borne BESS spectrometer flight in
August~2000~\cite{Shikaze2007}, close to solar maximum conditions.
The LIS specified in Eq.~(30) of Ref.~\cite{Bobik2012} has been used,
with parameters adopted from Fig.~3 of the reference. The tilt angles
have been extracted according to the ``L'' model as done in the
reference for the two cases discussed here.

The results are shown in Figs.~\ref{fig:helmod:ams98}
and~\ref{fig:helmod:bess00}, respectively. The \solarproptwo{}
implementation provides a decent description of the experimental data,
somewhat worse than the reference for the case of AMS-01 and somewhat
better at low energies for BESS-2000. Most importantly however, the
diffusion coefficients $\kappa_0$ have been scaled by a factor
\num{0.65} with respect to the values calculated from Eq.~(13) and
Table~1 of the reference for the results presented here. This large
correction underlines the inconsistency of the \solarproptwo{} results
with the calculations of Ref.~\cite{Bobik2012}. The flux predictions
obtained with the nominal values of $\kappa_0$ are shown by the curves
labeled ``\textit{no $\kappa$ scaling}''. They are much higher than
the measured fluxes and the reference results. This inconsistency
needs to be resolved, and this is one of the main motivations for the
publication of \solarproptwo{}.

Furthermore, according to the model description in
Ref.~\cite{Bobik2012}, the time step $\mathrm{d}t$ ``varies as''
$r^2/\kappa_{rr}$, whereas the \solarproptwo{} implementation uses a
default (fixed) time step of \qty{1000}{\second}. Taking the statement
at face value and setting $\mathrm{d}t=r^2/\kappa_{rr}$ leads to very
large time steps and consequently to a massive distortion of the fluxes
as illustrated in Figs.~\ref{fig:helmod:ams98}
and~\ref{fig:helmod:bess00}. In contrast, doubling the time step from
$\mathrm{d}t=\qty{1000}{\second}$ to $\mathrm{d}t=\qty{2000}{\second}$
does not change the results as seen in the figures.

In view of the differences to the reference results, two additional
tests of the reliability of the code base were performed. First, the
derivatives of the terms containing the diffusion coefficients in
eq.~(\ref{eq:sdes2d}) as well as $\vec{\nabla}\times(\vec{B}/B^2)$ in
eq.~(\ref{eq:helmod:drift}) were calculated by numerical
differentiation instead of the analytic expressions. The results are
identical except for small numerical differences (cf.~the curves
labeled ``\textit{Num.diff.}''). Second, the description of drifts
was simplified by setting $f(\theta)=1$ in eq.~(\ref{eq:helmod:drift})
and using eq.~(\ref{eq:simpledrift}) for the current sheet drift. As
seen by the curves labeled ``\textit{simple drifts}'', both
descriptions of drifts lead to very similar results.

To summarize, even though the model description in
Ref.~\cite{Bobik2012} is detailed and clear, there are again
differences in the results between two implementations of the same
model, calling for a detailed comparison of the underlying source
codes.

\section{Code performance}
\label{sec:code-performance}
\begin{figure}[tp]
  \centering
  \includegraphics[width=\textwidth]{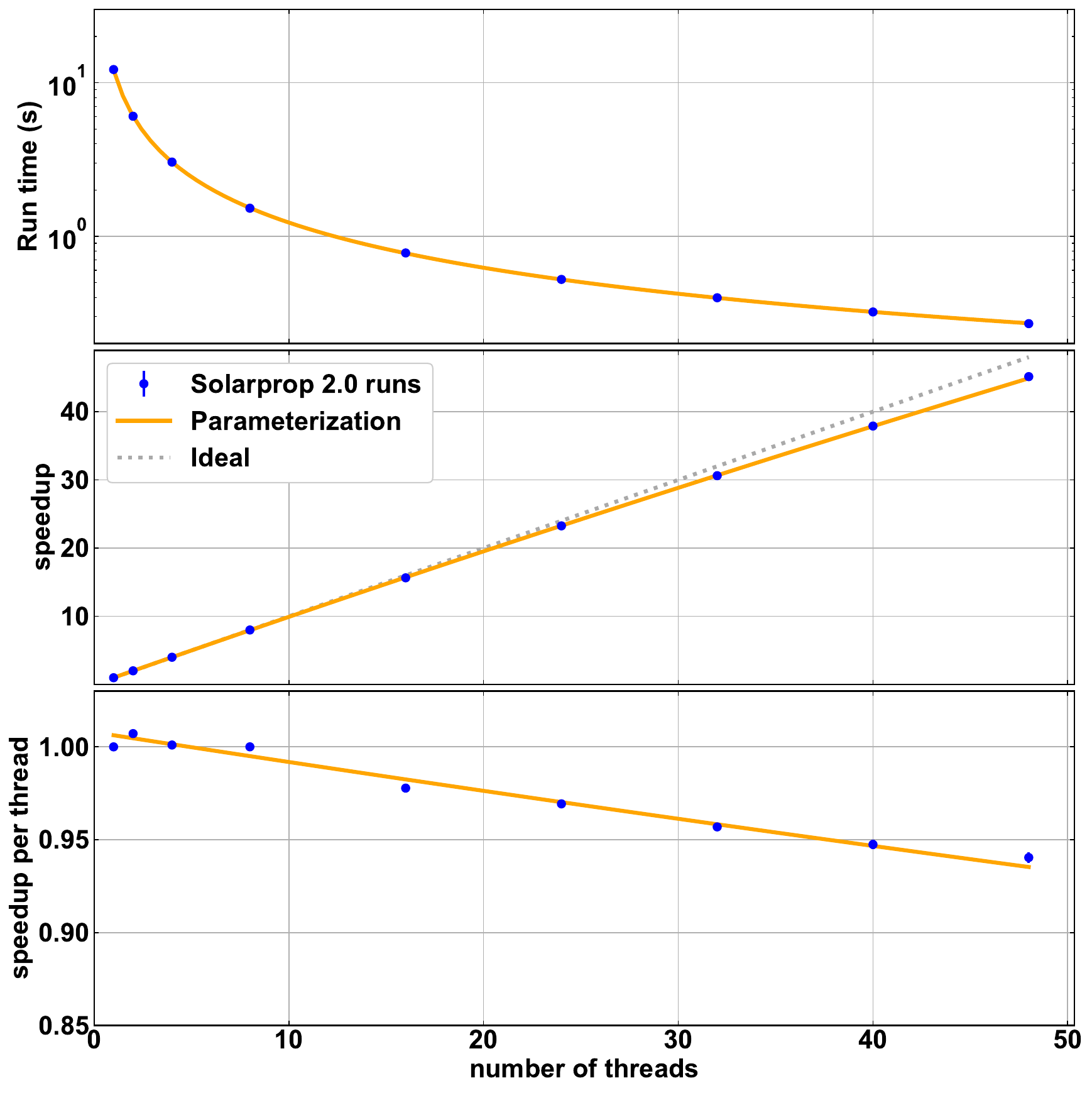}
  \caption{Scaling behavior of the \solarproptwo{} code. The overall
    run time (top), speedup (middle) and speedup per thread (bottom)
    are shown for a typical run. The ``Ideal'' case has $f_1=0$ in
    eq.~(\ref{eq:scaling}).}
  \label{fig:scaling}
\end{figure}
In \solarproptwo{}, the main computational loop, where the individual
pseudo-particles are propagated through the heliosphere as described
by the SDEs, is parallelized using OpenMP. The scaling behavior of the
code for a typical run is presented in Figure~\ref{fig:scaling}. The
top row shows the overall run time $t$ in seconds for different values
of the number of threads, up to a maximum of
48~threads. Assuming that the run time can be separated into a
single-threaded and a fully parallelized part, it can be parameterized
as
\begin{equation}
  \label{eq:scaling}
  t=t_0\left(f_1+\frac{1-f_1}{n_\mathrm{thr}}\right),
\end{equation}
where $n_\mathrm{thr}$ is the number of threads and $t_0$ is the run
time for single-thread execution, which depends on the concrete model
at hand. The fraction $f_1$ in Eq.~(\ref{eq:scaling}) represents the
time used for the part of the code that cannot be parallelized. The
parameterization drawn in Fig.~\ref{fig:scaling} uses a value of
$f_1=\num{1.6e-3}$. This indicates the excellent parallelization
achieved with \solarproptwo{}. The middle and bottom panels of
Fig.~\ref{fig:scaling} show the corresponding speedup and speedup per
thread, respectively. Note that the model of
Eq.~(\ref{eq:scaling}) is slightly optimistic. In reality, the
situation is more complex: the program has to wait until the last
pseudo-particle reaches the heliospheric boundary. To overcome this
problem, dynamic scheduling is used for the OpenMP loop, and
pseudo-particles of all energy bins are handled in one single
loop. This helps because the calculation is considerably faster for
particles at high energies than at low energies. So high-energy
particles can fill up the waiting times potentially caused by a
handful of low-energy particles with complicated pseudo-trajectories.

To put the performance of \solarproptwo{} into perspective, its run
time can be compared to that of the (closed-source) HelMod-4 code as
shown in Figure~5 of Ref.~\cite{Boschini2024}: for 1000
pseudo-particle trajectories of \qty{0.01}{GeV} protons during
Carrington Rotation 1937, the HelMod-4 code takes 20~minutes when run
on a CPU, while a GPU-accelerated version takes 2~minutes. By
contrast, the time it takes \solarproptwo{} to trace the same
particles in the model described in
sec.~\ref{sec:model-with-modified-mag-field}, which aims at the
re-implementation of an earlier version of HelMod, is close to
2~seconds even when only a single CPU core is used.


\section{Summary and outlook}
I have presented \solarproptwo{}, a greatly improved version of the
\solarprop{} code originally developed by Rolf Kappl for modeling
charge-sign dependent solar modulation of cosmic rays. The code
fulfills all design goals laid out in the introduction: it is
\emph{fast} due to many optimizations and, in particular, the
parallelization of the C++ backend with OpenMP, as demonstrated in
section~\ref{sec:code-performance}. It is \emph{accessible} due to
python bindings created with \texttt{pybind11} allowing users to set
up a model and calculate a modulated flux with a few lines of simple
\texttt{python} code. It is \emph{transparent} and \emph{versatile}
because the entire source code is available publicly online and the
code is structured such that new models can be added easily.

The code has been validated by reproducing several published results
by other groups, as detailed in section~\ref{sec:validation}. However,
there are also important discrepancies with two different
closed-source models which could not be resolved based on the
information found in the respective publications alone. Some hints at
potential problems in the reference implementations have been given,
but of course, it cannot be excluded that the discrepancies are caused
by errors in \solarprop. A detailed investigation on the level of the
source code will probably be necessary to resolve the issues. In any
case, the observed discrepancies highlight the need for open source
tools that can be used for independent cross-checks of model
calculations.

In the next step, it will then be possible to use \solarproptwo{} in
an effort of building a realistic model of solar modulation. The
experimental data available to guide this process are too numerous to
list here. But in particular, the AMS-02 experiment on the
International Space Station has been continuously collecting data for
fifteen years now, thus covering more than a solar
cycle. Time-dependent fluxes for cosmic-ray particles of different
masses and both positive and negative charges, namely protons and
antiprotons, positrons and electrons, have already been published for
a large fraction of this time period.

\section*{Acknowledgments}
I am deeply grateful to Rolf Kappl for creating \solarprop{} and
encouraging me to rework the original code, Stefan Schael for
discussions and comments on the draft manuscript, Benjamin Gudisch for
running the parallelization study, Andreas Kopp for answering my
questions about SDEs, and Stefano Della Torre for discussions about
HelMod. Experimental data have been extracted from the cosmic-ray
database CRDB~\cite{Maurin2014,Maurin2020}.

\appendix
\section{List of changes}
\label{sec:list-changes}
The following summarizes the list of changes and improvements in
\solarproptwo{} with respect to vanilla \solarprop. Please see the
\texttt{README.md} file for details.

\begin{itemize}
\item Add OpenMP support for parallelization of the main particle
  tracking loop. This results in a speedup close to the number of CPU
  cores.
\item Speedup by avoiding unnecessary calculations for particles in
  additional energy bins created by rebinning the original binning. A
  fine binning is needed for the histogram of particle energies at the
  edge of the heliosphere, to get a numerically accurate Green
  function matrix. But in vanilla Solarprop, pseudo-particles were
  also tracked for finer bins created from the binning of the
  top-of-the-atmosphere flux, even though the results of their
  tracking were never used. The fix results in a typical speedup
  factor of 5.
\item Add Python interface.
\item Switch to CMake build system.
\item Provide a script for downloading up-to-date data files from
  their respective sources.
\item Introduce a system of units, used throughout the C++ side.
\item Major code refactoring and restructuring. Rename models. Add
  factory class for models.
\item Generalize 2D SDEs to the case of $\kappa_{r\theta}\neq 0$. Add
  3D SDEs for models without polar diffusion,
  $\kappa_{r\theta}=\kappa_{\theta{}r}=\kappa_{\theta\phi}=\kappa_{\phi\theta}=0$.
\item New validation model for 3D SDEs: \texttt{Burger2012}.
\item Additional model implementations: \texttt{Strauss2012},
  \texttt{Helmod2012}, and \texttt{Helmod2018}.
\item Option for numerical calculation of derivatives needed for
  diffusion and drift terms.
\item Support for reading local interstellar input spectrum in FITS
  format was dropped on the C++ side. If needed, users can easily read
  FITS files on the python side.
\item Change SDEs to calculate
  $\mathrm{d}\cos\theta=-\sin\theta\,\mathrm{d}\theta$ instead of
  $\mathrm{d}\theta$, to avoid unphysical jumps in $\theta$ when close
  to zenith/nadir.
\item Switch SDEs to use $\mathrm{d}p$ instead of $\mathrm{d}T$, to
  avoid the extra loss term~\cite{Kopp2012,Bobik2016}.
\item Add scripts for visualizing pseudo-particle trajectories as well
  as the individual contributions to the transport process.
\item Add options for saving and loading the Green matrix. Allow
  subsequent modulation of different local interstellar spectra for
  the same model setup without re-running the SDE calculations.
\item Make sure that pseudo-particle trajectories end right at the
  heliospheric boundary, avoiding large (unphysical) energy changes
  for high-energy particles.
\item Use dynamic time steps: reduce step size close to the Sun in
  selected models.
\item Switch to completely new interpolation scheme. This avoids
  numerical problems at intermediate energies.
\item Various fixes and improvements for existing model
  implementations.
\end{itemize}

\section{Derivation of the full set of SDEs in two dimensions}
\label{app:sdes2d}

\noindent The derivation of the full set of SDEs in two dimensions
starts from Parker's equation (eq.~(\ref{eq:parker})),
\[
  \frac{\partial f}{\partial t}=\vec{\nabla}\cdot\left(\kappa\cdot\vec{\nabla}f-\vec{V}f\right)+\frac{1}{3p^2}(\vec{\nabla}\cdot\vec{V}_{sw})\frac{\partial p^3f}{\partial p},
\]
where $\vec{V}=\vec{V}_{sw}+\vec{V}_{dr}$. The individual
terms are written in 2D heliocentric spherical coordinates, and the
definitions $\vec{V}_{sw}=V_{sw}\,\vec{e}_r$,
$\vec{V}_{dr}\equiv V_{dr}\,\vec{e}_r + V_{d\theta}\,\vec{e}_\theta$,
and $\vec{\nabla}\cdot\vec{V}_{dr}=0$ are used:
\begin{equation}
  \begin{aligned}
    \vec{\nabla}\cdot\left(\kappa\cdot\vec{\nabla}f\right)
    &= \frac{1}{r^2}\frac{\partial}{\partial r}\left(r^2\kappa_{rr}\frac{\partial f}{\partial r}+r\kappa_{r\theta}\frac{\partial f}{\partial\theta}\right)\\
    &+ \frac{1}{r\sin\theta}\frac{\partial}{\partial\theta}\left(\sin\theta\kappa_{\theta r}\frac{\partial f}{\partial r}+\sin\theta\frac{\kappa_{\theta\theta}}{r}\frac{\partial f}{\partial\theta}\right),\\
  \end{aligned}
\end{equation}

\begin{equation}
  \begin{aligned}
    -\vec{\nabla}\cdot\vec{V}f & = -\vec{\nabla}\cdot\vec{V}_{sw}f-\vec{\nabla}\cdot\vec{V}_{dr}f\\
    & = -\frac{1}{r^2}\frac{\partial r^2 V_{sw}f}{\partial r}-f( \vec{\nabla}\cdot\vec{V}_{dr} ) - \vec{V}_{dr}\cdot\vec{\nabla}f\\
    & =  -\frac{1}{r^2}\frac{\partial(r^2 V_{sw})}{\partial r}f-V_{sw}\frac{\partial f}{\partial r}-V_{dr}\frac{\partial f}{\partial r}-\frac{V_{d\theta}}{r}\frac{\partial f}{\partial\theta},\\
  \end{aligned}
\end{equation}

\begin{equation}
  \begin{aligned}
    \frac{1}{3p^2}(\vec{\nabla}\cdot\vec{V}_{sw})\frac{\partial p^3f}{\partial p}&=\frac{1}{r^2}\frac{\partial (r^2V_{sw})}{\partial r}\frac{\partial p^3 f}{\partial p^3}\\
    &=\frac{1}{r^2}\frac{\partial (r^2V_{sw})}{\partial r}f + \frac{1}{r^2}\frac{\partial (r^2V_{sw})}{\partial r}p^3\frac{\partial f}{\partial p^3}\\
    &=\frac{1}{r^2}\frac{\partial (r^2V_{sw})}{\partial r}f + \frac{p}{3r^2}\frac{\partial (r^2V_{sw})}{\partial r}\frac{\partial f}{\partial p}.
  \end{aligned}
\end{equation}

\noindent Defining $\mu=\cos\theta$, which leads to
\begin{equation}
  \frac{1}{\sin\theta}\frac{\partial}{\partial\theta}=-\frac{\partial}{\partial\mu},
\end{equation}
the first term becomes
\begin{equation}
  \begin{aligned}
    \vec{\nabla}\cdot\left(\kappa\cdot\vec{\nabla}f\right)
    & = \frac{1}{r^2}\frac{\partial}{\partial r}\left(r^2\kappa_{rr}\frac{\partial f}{\partial r}-r\kappa_{r\theta}\sqrt{1-\mu^2}\frac{\partial f}{\partial\mu}\right)\\
    & + \frac{1}{r}\frac{\partial}{\partial\mu}\left(-\sqrt{1-\mu^2}\kappa_{\theta r}\frac{\partial f}{\partial r}+(1-\mu^2)\frac{\kappa_{\theta\theta}}{r}\frac{\partial f}{\partial\mu}\right)\\
    & = \kappa_{rr}\frac{\partial^2 f}{\partial r^2} + \frac{1}{r^2} \frac{\partial (r^2\kappa_{rr})}{\partial r}\frac{\partial f}{\partial r} \\
    & - \frac{\kappa_{r\theta}\sqrt{1-\mu^2}}{r}\frac{\partial^2 f}{\partial r\partial\mu} - \frac{\sqrt{1-\mu^2}}{r^2}\frac{\partial(r\kappa_{r\theta})}{\partial r}\frac{\partial f}{\partial\mu}\\
    & - \frac{\sqrt{1-\mu^2}\kappa_{\theta r}}{r}\frac{\partial^2 f}{\partial r\partial\mu}-\frac{1}{r}\frac{\partial(\sqrt{1-\mu^2}\kappa_{\theta r})}{\partial\mu}\frac{\partial f}{\partial r}\\
    & +\frac{(1-\mu^2)\kappa_{\theta\theta}}{r^2}\frac{\partial^2 f}{\partial\mu^2}+\frac{1}{r^2}\frac{\partial((1-\mu^2)\kappa_{\theta\theta})}{\partial\mu}\frac{\partial f}{\partial\mu}.\\
  \end{aligned}
\end{equation}
The complete transport equation becomes
\begin{equation}
  \begin{aligned}
    \frac{\partial f}{\partial t}
    & = \kappa_{rr}\frac{\partial^2 f}{\partial r^2} + \frac{1}{r^2} \frac{\partial (r^2\kappa_{rr})}{\partial r}\frac{\partial f}{\partial r}\\
    & - \frac{\kappa_{r\theta}\sqrt{1-\mu^2}}{r}\frac{\partial^2 f}{\partial r\partial\mu} - \frac{\sqrt{1-\mu^2}}{r^2}\frac{\partial(r\kappa_{r\theta})}{\partial r}\frac{\partial f}{\partial\mu}\\
    & - \frac{\sqrt{1-\mu^2}\kappa_{\theta r}}{r}\frac{\partial^2 f}{\partial r\partial\mu}-\frac{1}{r}\frac{\partial(\sqrt{1-\mu^2}\kappa_{\theta r})}{\partial\mu}\frac{\partial f}{\partial r}\\
    & +\frac{(1-\mu^2)\kappa_{\theta\theta}}{r^2}\frac{\partial^2 f}{\partial\mu^2}+\frac{1}{r^2}\frac{\partial((1-\mu^2)\kappa_{\theta\theta})}{\partial\mu}\frac{\partial f}{\partial\mu}\\
    & -V_{sw}\frac{\partial f}{\partial r}-V_{dr}\frac{\partial f}{\partial r}+\frac{\sqrt{1-\mu^2}\,V_{d\theta}}{r}\frac{\partial f}{\partial\mu}\\
    & + \frac{p}{3r^2}\frac{\partial (r^2V_{sw})}{\partial r}\frac{\partial f}{\partial p}.
  \end{aligned}
\end{equation}

\noindent Therefore, the transport equation has been expressed in the standard
form of the backward-in-time Kolmogorov equation,
\begin{equation}
  \label{eq:standardform}
  \frac{\partial f}{\partial t}=\sum\limits_iA_i(\vec{q},t)\frac{\partial f}{\partial q_i}+\frac{1}{2}\sum\limits_{i,j}\left[B(\vec{q},t)B^T(\vec{q},t)\right]_{i,j}\frac{\partial^2 f}{\partial q_i\partial q_j},
\end{equation}
with the corresponding SDEs given by
\begin{equation}
  \mathrm{d}\vec{q}=A(\vec{q},t)\,\mathrm{d}t+B(\vec{q},t)\,\mathrm{d}W,
\end{equation}
where $\mathrm{d}W$ is the increment of the Wiener process. With $\vec{q}=(r, \mu, p)$,
comparing the coefficients yields
\begin{equation}
  \begin{aligned}
  A_r &= \frac{1}{r^2}\frac{\partial}{\partial r}(r^2\kappa_{rr}) - \frac{1}{r}\frac{\partial}{\partial\mu}(\kappa_{r\theta}\sqrt{1-\mu^2}) - V_{sw} - V_{dr}, \\
  A_\mu &= -\frac{\sqrt{1-\mu^2}}{r^2}\frac{\partial}{\partial r}(r\kappa_{r\theta}) + \frac{1}{r^2}\frac{\partial}{\partial\mu}(\kappa_{\theta\theta}(1-\mu^2))+\frac{\sqrt{1-\mu^2}}{r}V_{d\theta}, \\
  A_p &= \frac{p}{3r^2}\frac{\partial r^2V_{sw}}{\partial r},\\
  \end{aligned}
\end{equation}
and
\begin{equation}
  \label{eq:bbt}
  BB^T=
  \begin{bmatrix}
    2\kappa_{rr} & -\frac{2\kappa_{r\theta}\sqrt{1-\mu^2}}{r} \\
    -\frac{2\kappa_{r\theta}\sqrt{1-\mu^2}}{r}& \frac{2\kappa_{\theta\theta}(1-\mu^2)}{r^2}\\
  \end{bmatrix}.
\end{equation}
The decomposition matrix $B$ is found to be
\begin{equation}
  \label{eq:matrixb}
  B=\sqrt{2}
  \begin{bmatrix}
    \frac{\sqrt{\kappa_{rr}\kappa_{\theta\theta}-\kappa_{r\theta}^2}}{\sqrt{\kappa_{\theta\theta}}} & -\frac{\kappa_{r\theta}}{\sqrt{\kappa_{\theta\theta}}} \\
    0 & \frac{\sqrt{1-\mu^2}\sqrt{\kappa_{\theta\theta}}}{r}
  \end{bmatrix}.
\end{equation}
Therefore the SDEs are given by
\begin{equation}
  \begin{aligned}
    \mathrm{d}r &= \left[\frac{1}{r^2}\frac{\partial}{\partial
        r}(r^2\kappa_{rr}) -
      \frac{1}{r}\frac{\partial}{\partial\mu}(\kappa_{r\theta}\sqrt{1-\mu^2})
      - V_{sw}-V_{dr}\right]\,\mathrm{d}t\\
    &+ w_r\,\frac{\sqrt{- 2 \kappa_{r \theta}^{2} + 2 \kappa_{rr}\kappa_{\theta \theta}}}{\sqrt{\kappa_{\theta \theta}}}\,\sqrt{\mathrm{d}t} - w_\mu \frac{\sqrt{2} \kappa_{r \theta}}{\sqrt{\kappa_{\theta \theta}}} \,\sqrt{\mathrm{d}t},\\
    \mathrm{d}\mu &= \left[
      -\frac{\sqrt{1-\mu^2}}{r^2}\frac{\partial}{\partial
        r}(r\kappa_{r\theta}) +
      \frac{1}{r^2}\frac{\partial}{\partial\mu}(\kappa_{\theta\theta}(1-\mu^2))+\frac{\sqrt{1-\mu^2}}{r}V_{d\theta}
    \right]\,\mathrm{d}t\\
    &+ w_\mu \frac{\sqrt{2(1 - \mu^{2})\kappa_{\theta \theta}}}{r} \,\sqrt{\mathrm{d}t},\\
    \mathrm{d}p & =\frac{p}{3r^2}\frac{\partial r^2V_{sw}}{\partial r}\,\mathrm{d}t,\\
  \end{aligned}
\end{equation}
where $w_r$ and $w_\mu$ are Gaussian random numbers with a mean of
zero and unit variance. A jupyter notebook containing this derivation
and the validation of the Cholesky decomposition of the matrix $BB^T$ is
found in the code repository.

\section{Derivation of the SDEs in three dimensions without polar diffusion}
\label{app:sdes3d}

\noindent The derivation of the SDEs in three dimensions starts again
from Parker's equation.  In 3D spherical coordinates, and using the
definitions $\vec{V}_{sw}=V_{sw}\,\vec{e}_r$,
$\vec{V}_{dr}\equiv V_{dr}\,\vec{e}_r + V_{d\theta}\,\vec{e}_\theta +
V_{d\phi}\,\vec{e}_\phi$ , and $\vec{\nabla}\cdot\vec{V}_{dr}=0$, the
individual terms are:
\begin{equation}
  \begin{aligned}
    \vec{\nabla}\cdot\left(\kappa\cdot\vec{\nabla}f\right)&=\frac{1}{r^2}\frac{\partial}{\partial r}\left(r^2\kappa_{rr}\frac{\partial f}{\partial r}+r\kappa_{r\theta}\frac{\partial f}{\partial\theta}+\frac{r\kappa_{r\phi}}{\sin\theta}\frac{\partial f}{\partial\phi}\right)\\
    &+\frac{1}{r\sin\theta}\frac{\partial}{\partial\theta}\left(\sin\theta\kappa_{\theta r}\frac{\partial f}{\partial r}+\sin\theta\frac{\kappa_{\theta\theta}}{r}\frac{\partial f}{\partial\theta}+\frac{\kappa_{\theta\phi}}{r}\frac{\partial f}{\partial\phi}\right)\\
    &+\frac{1}{r\sin\theta}\frac{\partial}{\partial\phi}\left(\kappa_{\phi{}r}\frac{\partial f}{\partial{}r} + \frac{\kappa_{\phi\theta}}{r}\frac{\partial f}{\partial\theta} + \frac{\kappa_{\phi\phi}}{r\sin\theta}\frac{\partial f}{\partial\phi}\right),\\
  \end{aligned}
\end{equation}

\begin{equation}
  \begin{aligned}
    -\vec{\nabla}\cdot\vec{V}f &= -\vec{\nabla}\cdot\vec{V}_{sw}f-\vec{\nabla}\cdot\vec{V}_{dr}f \\
    & = -\frac{1}{r^2}\frac{\partial r^2 V_{sw}f}{\partial r}-f( \vec{\nabla}\cdot\vec{V}_{dr} ) - \vec{V}_{dr}\cdot\vec{\nabla}f \\
    & =  -\frac{1}{r^2}\frac{\partial(r^2 V_{sw})}{\partial r}f-V_{sw}\frac{\partial f}{\partial r}-V_{dr}\frac{\partial f}{\partial r}-\frac{V_{d\theta}}{r}\frac{\partial f}{\partial\theta}-\frac{V_{d\phi}}{r\sin\theta}\frac{\partial f}{\partial\phi},\\
  \end{aligned}
\end{equation}

\begin{equation}
  \begin{aligned}
\frac{1}{3p^2}(\vec{\nabla}\cdot\vec{V}_{sw})\frac{\partial p^3f}{\partial p}&=\frac{1}{r^2}\frac{\partial (r^2V_{sw})}{\partial r}\frac{\partial p^3 f}{\partial p^3}\\
&=\frac{1}{r^2}\frac{\partial (r^2V_{sw})}{\partial r}f + \frac{1}{r^2}\frac{\partial (r^2V_{sw})}{\partial r}p^3\frac{\partial f}{\partial p^3}\\
&=\frac{1}{r^2}\frac{\partial (r^2V_{sw})}{\partial r}f + \frac{p}{3r^2}\frac{\partial (r^2V_{sw})}{\partial r}\frac{\partial f}{\partial p}.
  \end{aligned}
\end{equation}

\noindent Assuming the case of no polar diffusion,
\begin{equation}
  \kappa_{r\theta}=\kappa_{\theta{}r}=\kappa_{\theta\phi}=\kappa_{\phi\theta}=0,
\end{equation}
and using the definition $\mu=\cos\theta$, the transport equation
becomes
\begin{equation}
  \begin{aligned}
    \frac{\partial f}{\partial t}&=\kappa_{rr}\frac{\partial^2 f}{\partial r^2} + \frac{1}{r^2} \frac{\partial (r^2\kappa_{rr})}{\partial r}\frac{\partial f}{\partial r} + 
    \frac{\kappa_{r\phi}}{r\sqrt{1-\mu^2}}\frac{\partial^2f}{\partial{}r\partial\phi}+\frac{1}{r^2\sqrt{1-\mu^2}}\frac{\partial(r\kappa_{r\phi})}{\partial{}r}\frac{\partial f}{\partial\phi}\\
    &+\frac{(1-\mu^2)\kappa_{\theta\theta}}{r^2}\frac{\partial^2 f}{\partial\mu^2}+\frac{1}{r^2}\frac{\partial((1-\mu^2)\kappa_{\theta\theta})}{\partial\mu}\frac{\partial f}{\partial\mu} +
    \frac{1}{r\sqrt{1-\mu^2}} \frac{\partial\kappa_{\phi r}}{\partial\phi} \frac{\partial f}{\partial r} + 
    \frac{\kappa_{\phi r}}{r\sqrt{1-\mu^2}} \frac{\partial^2 f}{\partial\phi\partial r}\\
    &+\frac{\kappa_{\phi\phi}}{r^2(1-\mu^2)} \frac{\partial^2 f}{\partial\phi^2}+ 
    \frac{1}{r^2(1-\mu^2)} \frac{\partial\kappa_{\phi\phi}}{\partial\phi} \frac{\partial f}{\partial\phi}\\
    &-V_{sw}\frac{\partial f}{\partial r}-V_{dr}\frac{\partial f}{\partial r}+\frac{\sqrt{1-\mu^2}V_{d\theta}}{r}\frac{\partial f}{\partial\mu}-\frac{V_{d\phi}}{r\sqrt{1-\mu^2}}\frac{\partial f}{\partial\phi} +
    \frac{p}{3r^2}\frac{\partial (r^2V_{sw})}{\partial r}\frac{\partial f}{\partial p},\\
  \end{aligned}
\end{equation}
which is again the standard form of the backward-in-time Kolmogorov
equation, eq.~(\ref{eq:standardform}). With
$\vec{q}=(r, \mu, \phi, p)$, comparing the coefficients yields
\begin{equation}
  \begin{aligned}
    A_r&=\frac{1}{r^2}\frac{\partial}{\partial r}(r^2\kappa_{rr}) + \frac{1}{r\sqrt{1-\mu^2}} \frac{\partial\kappa_{r\phi}}{\partial\phi} - V_{sw} - V_{dr},\\
    A_\mu&=\frac{1}{r^2}\frac{\partial}{\partial\mu}(\kappa_{\theta\theta}(1-\mu^2))+\frac{\sqrt{1-\mu^2}V_{d\theta}}{r},\\
    A_\phi&=\frac{1}{r^2\sqrt{1-\mu^2}}\frac{\partial(r\kappa_{r\phi})}{\partial{}r} + \frac{1}{r^2(1-\mu^2)} \frac{\partial\kappa_{\phi\phi}}{\partial\phi} -\frac{V_{d\phi}}{r\sqrt{1-\mu^2}},\\
    A_p&=\frac{p}{3r^2}\frac{\partial r^2V_{sw}}{\partial r},\\
  \end{aligned}
\end{equation}
and
\begin{equation}
  \begin{aligned}
    BB^T=
    \begin{bmatrix}
      2\kappa_{rr} & 0 & 2\frac{\kappa_{r\phi}}{r\sqrt{1-\mu^2}}\\
      0 & \frac{2(1-\mu^2)\kappa_{\theta\theta}}{r^2} & 0 \\
      2\frac{\kappa_{r\phi}}{r\sqrt{1-\mu^2}}& 0 & 2\frac{\kappa_{\phi\phi}}{r^2(1-\mu^2)}\\
    \end{bmatrix}.
  \end{aligned}
\end{equation}
The decomposition matrix $B$ is found to be
\begin{equation}
   B=\sqrt{2}
   \begin{bmatrix}
      \sqrt{\kappa_{rr} - \frac{\kappa_{r \phi}^{2}}{\kappa_{\phi \phi}}} & 0 & \frac{\kappa_{r \phi}}{\sqrt{\kappa_{\phi \phi}}}\\
      0 & \frac{\sqrt{1-\mu^{2}} \sqrt{\kappa_{\theta \theta}}}{r} & 0\\
      0 & 0 & \frac{\sqrt{\kappa_{\phi \phi}}}{r \sqrt{1 - \mu^{2}}}
    \end{bmatrix}.
\end{equation}
Therefore the SDEs are given by
\begin{equation}
  \begin{aligned}
    \mathrm{d}r&=\left[\frac{1}{r^2}\frac{\partial}{\partial r}(r^2\kappa_{rr}) + \frac{1}{r\sqrt{1-\mu^2}} \frac{\partial\kappa_{r\phi}}{\partial\phi} - V_{sw} - V_{dr}\right]\,\mathrm{d}t\\
    &+w_r\, \sqrt{2\kappa_{rr}-\frac{2\kappa_{r\phi}^2}{\kappa_{\phi\phi}}} \,\sqrt{\mathrm{d}t} + w_\phi \frac{\sqrt{2} \kappa_{r \phi}}{\sqrt{\kappa_{\phi\phi}}} \,\sqrt{\mathrm{d}t},\\
    \mathrm{d}\mu&=\left[ \frac{1}{r^2}\frac{\partial}{\partial\mu}(\kappa_{\theta\theta}(1-\mu^2))+\frac{\sqrt{1-\mu^2}V_{d\theta}}{r} \right]\,\mathrm{d}t\\
    &+w_\mu \frac{\sqrt{2(1-\mu^2)\kappa_{\theta\theta}}}{r} \,\sqrt{\mathrm{d}t},\\
    \mathrm{d}\phi&=\left[ \frac{1}{r^2\sqrt{1-\mu^2}}\frac{\partial(r\kappa_{r\phi})}{\partial{}r} + \frac{1}{r^2(1-\mu^2)} \frac{\partial\kappa_{\phi\phi}}{\partial\phi} -\frac{V_{d\phi}}{r\sqrt{1-\mu^2}} \right]\,\mathrm{d}t\\
    &+w_\phi \frac{\sqrt{2\kappa_{\phi\phi}}}{r\sqrt{1-\mu^2}} \,\sqrt{\mathrm{d}t},\\
    \mathrm{d}p&=\frac{p}{3r^2}\frac{\partial r^2V_{sw}}{\partial r}\,\mathrm{d}t,\\
  \end{aligned}
\end{equation}
where $w_r$, $w_\mu$, and $w_\phi$ are Gaussian random numbers with a
mean of zero and unit variance. Again, a jupyter notebook with
additional details on the derivation is found in the code repository.

\section{Nomenclature}
Tables~\ref{tab:symbols:1} and~\ref{tab:symbols:2} contain the list of
symbols used throughout this paper and also in the documentation of
the source code and the associated jupyter notebooks.
\begin{table}[hp]
  \centering
  \begin{tabular}{p{0.15\linewidth} p{0.8\linewidth}}
    $A$ & polarity of the heliospheric magnetic field: $A=\pm1$ \\
    $\alpha$ & tilt angle of the heliospheric current sheet \\
    $\vec{B}$ & heliospheric magnetic field vector \\
    $B_0$ & scale factor determining the strength of the heliospheric
            magnetic field \\
    $B_\oplus$ & amplitude of the heliospheric magnetic field at Earth \\
    $\beta$ & $v/c$, particle velocity divided by the speed of light \\
    $\beta^\prime$ & local inclination of the current sheet with
                     respect to the radial direction \\
    $c$ & speed of light \\
    $d$ & distance to the point of closest approach on the current sheet \\
    $\delta(\theta)$ & latitudinal correction function to magnetic field \\
    $\delta_0$ & scale factor for latitudinal correction function to
                 magnetic field \\
    $\mathrm{d}r$ & SDE pseudo-particle step in radial direction \\
    $\mathrm{d}\mu$ & SDE pseudo-particle step in $\cos\theta$ \\
    $\mathrm{d}\phi$ & SDE pseudo-particle step in azimuthal direction \\
    $\mathrm{d}t$ & SDE increment in backward-time \\
    $\vec{e}_r$ & unit vector in radial direction \\
    $\vec{e}_\theta$ & unit vector in polar direction \\
    $\vec{e}_\phi$ & unit vector in azimuthal direction \\
    $f$ & omnidirectional phase-space distribution function \\
    $\tilde{f}$ & transition function used for the effect of the
                  current sheet \\
    $\Gamma$ & defined as $\Omega{}r\sin\theta/V_{sw}$ \\
    $H$ & Heaviside step function \\
    $\iota$ & ratio $\kappa_{\perp{}\theta}/\kappa_{\perp{}r}$ \\
    $K(R)$ & rigidity dependence of the diffusion coefficient \\
    $k$ & rigidity-dependent scale used in the \texttt{Burger2012}
          model for the reach of the current sheet \\
    $\kappa$ & diffusion coefficient matrix \\
    $\kappa_{rr}$, $\kappa_{\theta\theta}$, $\kappa_{\phi\phi}$, $\kappa_{r\theta}$, $\kappa_{r\phi}$, $\kappa_{\theta\phi}$ &
       components of the diffusion tensor in heliospheric spherical coordinates \\
    $\kappa_0$ & scale factor determining the overall diffusion
                 coefficient scale \\
    $\kappa_\parallel$ & diffusion coefficient in field aligned
                         coordinates, parallel to the magnetic field \\
    $\kappa_\perp$ & diffusion coefficient in field aligned
                     coordinates, perpendicular to the magnetic field \\
    $\kappa_{\perp{}r}$ & diffusion coefficient in field aligned
                          coordinates, perpendicular to the magnetic field in radial direction \\
    $\kappa_{\perp{}\theta}$ & diffusion coefficient in field aligned
                               coordinates, perpendicular to the magnetic field in polar direction \\
    $\lambda_\parallel$ & parallel mean free path \\
    $\mu$ & $\cos\theta$ \\
  \end{tabular}
  \caption{List of symbols (part 1).}
  \label{tab:symbols:1}
\end{table}

\begin{table}[hp]
  \centering
  \begin{tabular}{p{0.15\linewidth} p{0.8\linewidth}}
    $N_\mathrm{sunspot}$ & 13-month smoothed monthly total sunspot
                           number \\
    $\nu$ & angle between the normal to the current sheet and the $-\vec{e}_\theta$
            direction \\
    $\Omega$ & average angular velocity of the Sun \\
    $p$ & particle momentum \\
    $\phi$ & azimuthal heliospheric coordinate \\
    $\varphi$ & modulation potential \\
    $\Phi$ & differential particle flux \\
    $\psi$ & spiral angle, defined via
             $\tan\psi=-B_\phi/\sqrt{B_r^2+B_\theta^2}$ \\
    $q$ & particle charge \\
    $r$ & radial heliospheric coordinate \\
    $r_0$ & mean distance of Earth to the Sun \\
    $r_L$ & Larmor radius $R/B$ of the particle \\
    $r_\odot$ & solar radius \\
    $R$ & rigidity $p/q$ of the particle \\
    $R_h$ & radius of the heliosphere \\
    $\rho$ & ratio $\kappa_{\perp{}r}/\kappa_\parallel$ \\
    $t$ & time coordinate \\
    $T$ & kinetic energy \\
    $\theta$ & colatitude: polar heliospheric coordinate \\
    $\theta^\prime$ & polar angle determining the location of the
                      heliospheric current sheet \\
    $v$ & particle velocity \\
    $\vec{V}_{dr}$ & sum of the pitch-angle averaged guiding center drift
                     velocity in the heliospheric magnetic field and
                     the drift velocity along the heliospheric current sheet \\
    $\vec{v}_\mathrm{hcs}$ & drift velocity along the heliospheric
                             current sheet \\
    $V_{dr}$ & radial component of the drift velocity \\
    $V_{d\theta}$ & polar component of the drift velocity \\
    $V_{d\phi}$ & azimuthal component of the drift velocity \\
    $\vec{V}_{sw}$ & solar wind speed \\
    $\vec{V}$ & $\vec{V}_{sw}+\vec{V}_{dr}$ \\
    $v_\mathrm{omni}$ & 27-day averaged value of solar wind speed as
                        extracted from \texttt{OMNIweb} \\
    $w_r$, $w_\mu$, $w_\phi$ & Gaussian random numbers with zero mean
                               and unit variance \\
    $\xi$ & angle defined via $\tan\xi=B_\theta/B_r$ \\
  \end{tabular}
  \caption{List of symbols (part 2).}
  \label{tab:symbols:2}
\end{table}

\clearpage

\bibliographystyle{elsarticle-num} 
\bibliography{sprop2_papers}

\end{document}